\documentclass[final,3p,times]{elsarticle}

\usepackage{lineno,hyperref}
\modulolinenumbers[5]

\journal{Icarus}

\biboptions{authoryear}

\begin{document}

\begin{frontmatter}

\title{A recent origin for Saturn's rings from the collisional disruption of an icy moon}

\author{John Dubinski\fnref{myfootnote}}
\address{Canadian Institute for Theoretical Astrophysics, 60 St. George
St., University of Toronto, Toronto, Ontario, Canada M5S 3H8}
\fntext[myfootnote]{e-mail: dubinski@cita.utoronto.ca}


\begin{abstract}
The disruption of an icy moon in a collision with an interloping comet 
a few hundred million years ago is a simple way to create Saturn's rings.
A ring parent moon with a mass comparable to Mimas
could be trapped in mean motion resonance with Enceladus and Dione
in an orbit near the current outer edge of the rings just beyond
the Roche zone.
I present collisional N-body simulations of
cometary impacts that lead to the partial disruption of 
a differentiated moon with a rocky core and icy mantle.
The core can survive largely intact while the debris from the mantle 
settles into a ring of predominantly ice particles straddling the orbital 
radius of the parent moon.  
The nascent ring spreads radially due to collisional viscosity
while mass re-accretes onto the remnant rocky core to form a new 
moon that can be identified as Mimas.  
The icy debris that migrates into the Roche zone
evolves into Saturn's ring system.  
Torques from tidal interaction with Saturn and resonant interactions with the 
rings push the recently formed Mimas outward to its current position on the
same timescale of a few hundred million years.
This scenario accounts for the high ice fraction observed in Saturn's
rings and explains why the ring mass is comparable to the mass of Mimas.
The prior existence of a ring parent moon in mean motion resonance 
results in a tidal heating rate for Enceladus in the recent past that is significantly 
larger than the current rate.
\end{abstract}

\begin{keyword}
Saturn, rings, satellites; Satellite, dynamics; Enceladus; Mimas
\end{keyword}

\end{frontmatter}

%\linenumbers
\newcommand{\dd}{\mathrm{d}}
\newcommand{\pc}{\mathrm{pc}}
\newcommand{\mpc}{\mathrm{Mpc}}
\newcommand{\msun}{M_{\odot}}
\newcommand{\be}{\begin{equation}}
\newcommand{\ee}{\end{equation}}
\newcommand{\bea}{\begin{eqnarray}}
\newcommand{\eea}{\end{eqnarray}}
\newcommand{\appgeq}{\stackrel{>}{\sim}}
\newcommand{\appleq}{\stackrel{<}{\sim}}
\newcommand{\ec}{{\cal E}}


%%%%%%%%%%%%%%%%%%%%%%%%%%%%%%%%%%%%%%%%%%%%%%%%%%

%%%%%%%%%%%%%%%%% BODY OF PAPER %%%%%%%%%%%%%%%%%%

\section{Introduction}

The origin of Saturn's rings has been an ongoing puzzle since Galileo's
discovery of the system four hundred years ago.
Current theories that explain the rings' origin 
are constrained by the mass, composition and age of the rings
\citep{harris84,dones91,charnoz09a,canup10,charnoz09b,cuzzi10} (See \citet{charnoz18} for a comprehensive review of the origin of planetary rings.)
Observations of density waves in the A  and B rings and comparisons of
simulations of optical depth to Cassini observations for the denser B ring find a total mass of the rings
comparable to Saturn's moon Mimas \citep{esposito83,robbins10,hedman16} with uncertainties due to clumpiness of matter in the rings
leading to a range varying from $2-10 \times 10^{19}\;{\rm kg}$ 
cf. $M_{Mimas}=3.75\times 10^{19}\;{\rm
kg}$.\footnote{At the Cassini Science Symposium in August 2018, Luciano Iess presented results giving
estimates for the total mass of the A, B, and C rings based on
Cassini close flybys as $0.41\pm 0.04 M_{Mimas}$ or $1.5\times
10^{19}\;{\rm kg}$ putting it at the low end of the range of previous estimates;
\url{http://lasp.colorado.edu/media/projects/cassini/meetings/css2018/4-THURSDAY/3-Saturn_1_(S1-)/6-Iess-REVISED.pdf}}
%\footnote{A news item on the American Geophysical Meeting in December 2017 quotes Luciano Iess giving 
%estimates for the B ring mass based on
%the Cassini close flybys of the rings as 0.4 $M_{Mimas}$ or $0.15\times
%10^{20}\;{\rm kg}$ putting it at the low end of the previous estimates;
%\url{http://www.sciencemag.org/news/2017/12/saturn-s-rings-are-recent-addition-solar-system-cassini-observations-show}}
The rings are mainly composed of ice with historical radio and radar observations suggesting an ice mass fraction greater
than 90\% \citep{cuzzi10}.  
The latest microwave observations from Cassini and the VLA imply ice mass fractions 
greater than $99\%$ for the A and B rings and slightly less for the C ring \citep{zhang17a,zhang17b,zhang19} 
These measurements are inconsistent with 
the mass-weighted average value of 60\% ice/40\% rock
for Saturn's 5 classical inner moons \citep{matson09}.   If 
the rings and moons form coevally in Saturn's early history 
and are born from a massive spreading ring
\citep{charnoz09b,crida12},
one needs to explain their different compositions.
One proposed solution to this problem
is a scenario in which a differentiated moon
with a mass comparable to Titan falls into Saturn at early times and
loses its icy mantle via tidal stripping to form a nearly pure ice 
ring \citep{canup10}.
The final constraint on ring origin theories comes from the age of 
the rings.  Two arguments suggest that the rings may be as young as a few
hundred million years causing problems for theories requiring a primordial origin.
The viscosity arising from dissipative particle collisions implies
that a ring will spread to conserve angular momentum while losing orbital 
energy \citep{goldreichtremaine82}.
The viscous timescale is $t_{\nu} \sim \Delta R^2/\nu$ where $\Delta R$ is 
the current radial width of the rings 
and $\nu$ is the viscosity with the underlying assumption that the ring originates
as a radially thin object.  
This timescale is a
few hundred million years based on the current estimates of ring viscosity 
for the A ring at least \citep{esposito86,daisaka01}.
The more massive B ring could be much older based on similar arguments
and more detailed treatments of the sources of viscosity in the denser B ring 
may imply longer evolution and survival times with the rings perhaps as old as the age of the Solar System 
supporting primordial origin theories \citep{daisaka01,salmon10}.
Meteoroid bombardment can also change the 
composition of the rings over time through contamination from
mixing with non-icy constituents.  Modeling of the spectral evolution of the
rings under exposure to meteoroids also implies a young age of only
a few hundred million years again \citep{durisen92,cuzzi98,estrada15} while new results from Cassini \citep[e.g.,][]{zhang17b} imply exposure ages in the range of 30-150 Myr.
These results are confounding to primordial ring origin theories.
Scenarios where the rings might form recently include 
collisional disruption of a moon or moons within the Roche zone through impacts with infalling comets \citep{harris84} and
tidal stripping of a massive icy Kuiper belt object passing close to Saturn \citep{dones91,hyodo17a}. 
Both scenarios can happen at any time but would most likely occur during
the Late Heavy Bombardment \citep{charnoz09b}.
The collisional disruption theory is problematic since it is hard to form or move moons near the Roche zone as well as hold them in
place due to tidal evolution.  
The tidal stripping theory 
seems unlikely to have occurred in recent times because of the requirement of a large Kuiper belt object - perhaps as massive as Pluto -  having a near
collision with Saturn.  
A complete collisional disruption requires a parent object that is nearly pure ice - a fact inconsistent with the composition of the icy
moons.  However, in a tidal disruption model the icy mantle of a differentiated parent object might be stripped to create an icy ring \citep{hyodo17a}.
The recent determination of small tidal dissipation factor $Q$ for Saturn \citep{lainey12,lainey17} implies another possible scenario where
orbital instabilities can be driven by the more rapid tidal evolution of the icy satellite system.  This could lead to 
moon-moon collisions within
the past $10^8$ years that formed the current icy moon system and rings \citep{cuk16}.  \citet{hyodo17b} have explored this scenario and show
that the debris from a hypothetical collision between a proto-Rhea and proto-Dione at $a=500000\;{\rm km }$ quickly reaccretes into other moons
before spreading into the Roche zone making this scenario unlikely as a way to form the rings though more work needs to be done to understand this process.
In summary, the inferred mass and composition of the rings create tension between theories 
that propose a primordial and recent origin and there is no strong consensus
on a formation mechanism.

In this paper, we propose a modified version of the collisional disruption mechanism
of ring formation \citep{harris84} that is consistent with the mass and composition of the rings.
In this model, a comet on a heliocentric orbit - an Oort-cloud comet or Centaur - with
sufficient mass and impact energy hits and disrupts a ring parent moon.
The resulting 
debris then spreads along the moon's orbit to form the ring system.
%\footnote{Perhaps a suitable name for this moon is Menoetius: a
%Titan, son of Iapetus and brother of Prometheus, struck down by Zeus with a thunderbolt for his hubris and cast into 
%the underworld during the battle of the Titans (Hesiod, Theogony, 507).}
This mechanism is likely relevant to the formation of the less massive rings 
of Jupiter, Uranus and Neptune but faces difficulties in making Saturn's rings \citep{colwell94} because of their relatively large mass and
high ice fraction.
The composition of a hypothetical ring parent moon for Saturn should be consistent
with those of Saturn's moons shown in Table \ref{table:moons} \citep{matson09,thomas10}.
\begin{table*}
\centering
\begin{tabular}{rrrrr}
\hline
Object & a ($10^5 km$) & Mass ($10^{20}\;{\rm kg}$) & $\bar{\rho}$ (${\rm g/cm^3}$) & Ice Fraction\\
\hline
A Ring & 1.22-1.37 & 0.06 & - & 90-95\% \\
B Ring & 0.92-1.18 & 0.2-1.0 & - & 90-95\% \\
Janus &  1.51 & 0.020 & 0.63 & - \\
Mimas  & 1.85 & 0.375 & 1.15 & 74\% \\
Enceladus & 2.38 & 1.08 & 1.61 & 43\% \\
Tethys & 2.95 & 6.17 & 0.99 & 94\% \\
Dione & 3.77 & 10.95 & 1.48 & 50\% \\
Rhea & 5.27 & 23.07 & 1.24 & 67\% \\
\hline
Ring parent & 1.40 & 0.750 & 1.03 & 84\% \\
\hline
\end{tabular}
\caption{The properties of Saturn's rings \citep{robbins10,cuzzi10,hedman16} and mid-sized moons \citep{matson09,thomas10} compared 
to a hypothetical ring parent moon.
Theories based on a coeval origin of the rings and moons have difficulty in explaining a high 
ice mass fraction in the rings compared to the moons.  The properties of the proposed ring parent moon are included for comparison.
} 
\label{table:moons}
\end{table*}
Since the mass of the rings is comparable to the mass of Saturn's moon Mimas \citep{robbins10},
the disrupted moon must be at least this large and one further needs to find a way to
create a ring made of nearly pure ice.
The ice fractions of Saturn's 5 mid-sized moons vary from
43\% (Enceladus) to 94\% (Tethys) with an mass-weighted 
average value of about 60\% \citep{thomas10} (see Table \ref{table:moons}).
The complete disruption of a moon with the current average composition of the extant moons
would create a ring inconsistent with the observations
though perhaps a parent moon with the composition of Tethys might be adequate.

A further problem with this collisional scenario arises when
considering the initial orbital radius of 
a hypothetical ring parent moon.   
To remain stable in the presence of Saturn's tidal forces, 
the parent moon must be located outside of 
the \citet{roche1847} radius $R_{Roche}=2.45 R_S (\rho_S/\rho_{moon})^{1/3}$
where $R_S$ and $\rho_S$ are the radius and mean density of Saturn and $\rho_{moon}$ is the moon's density.
However, the parent moon cannot be too far removed from the Roche radius so
that some of the collisional debris can migrate inwards to form the 
rings before reassembling into a new moon.
For a moon composed of pure ice, $R_{Roche}\approx 130000\;{\rm km}$ consistent with the 
rings current radial extent of $R<137000\;{\rm km}$,  so the parent moon 
should ideally located near this radius. 
The co-orbital moons Janus and Epimetheus are already near the current edge of the rings but their collisional disruption
would create a ring with a mass that is an order of magnitude too small. 
Assuming the disrupting parent moon is massive enough,
the debris would settle into a ring promptly and then spread radially
on the viscous timescale discussed above.
If the disruption occurred recently - say within the past 500 Myr, the primordial existence of a Mimas-sized moon located 
just near the Roche radius might seem unlikely since tidal friction \citep[e.g.,][]{goldreich66} would have caused a moon
of this mass to migrate away from this position to Mimas's current orbital radius within the age of the solar system 
based on the conventional values of the 
Saturn's Love number $k_{2,S}$ and tidal dissipation function $Q_S$ having a ratio $k_{2,S}/Q_S=2.0\times 10^{-5}$.  
Recent analysis of a century of astrodynamical measurements combined with Cassini's observations
of Saturn's moons imply values $k_{2,S}/Q_S\approx 1.6\times 10^{-4}$ reducing 
the tidal evolution timescale by an order of magnitude making it even more difficult
to consider a moon existing at this position for long \citep{lainey12,lainey17} and in fact
implying short timescales for potential catastrophic orbital instabilities for the entire inner moon system \citep{cuk16}.
However, \citet{fuller16} have argued that Saturn's effective $Q$ may be intermittently diminished 
by resonant locking between dynamical tidal oscillations commensurate with orbital frequencies of satellites
while the long term average of $Q_S$ would be closer to the conventional value.
The implications of a evolving value of $Q_S$ have not been fully explored.  
%For this paper, we will assume
%the lower conventional value for the tidal ratio $k_{2,S}/Q_S$.
The only way to hold a moon in place near the Roche radius over the age of the solar system
prior to a recent disruption is via an
$e$-type mean motion resonant (MMR) trapping with the other icy moons. 
Another difficulty is that such destructive events may be rare with the disruption
of a ring parent moon of the mass of Mimas happening once every 20 Gyr at the current
cometary flux rate giving a probability of destruction
in the last 500 Myr of $\sim 2\%$ \citep{lissauer88,dones09,charnoz09a}.
The rarity of an event in itself does not necessarily rule out its occurrence if no alternative 
more probable mechanisms can be found.

The scenario proposed in this paper overcomes the above difficulties 
and provides a modified collisional scenario for a recent origin of Saturn's rings.
The strange fact that the mass of the rings is close to the
mass of Mimas may not be a coincidence but rather a clue pointing to a collisional disruption 
origin.\footnote{It should be noted that primordial origin theories that posit more complex viscosity evolution 
for the rings also result in a ring with a mass comparable to Mimas \citep{salmon10}.}
Now consider the collisional disruption of a ring parent moon 
that is twice the mass of Mimas located 
just outside the current edge of the rings near the Roche radius
at a radius $a \approx 140000\;{\rm km}$. 
After disruption, half of the collisional
debris migrates inwards within the Roche zone to become a new ring while 
the remainder of the debris migrates outwards
along with any remnant of the collision and likely reassembles to become a new moon following the mechanism
described by \citet{crida12}.
One might identify this new moon with Mimas itself. 
In this picture, Mimas forms coevally with the rings and migrates to its
current position from the time the rings formed.  

To support this hypothesis, imagine taking the total mass of
Saturn's A and B rings and Mimas itself and placing it within a single body - a hypothetical ring parent moon 
composed of mass from the rings and Mimas.
Using reasonable estimates of the ring surface density \citep[e.g.,][]{robbins10}
and assuming there is no significant mass loss from the Saturn system during the disruption, 
one can calculate the total mass and angular
momentum of this combined system and compute the orbital radius $a_{parent}$ for this hypothetical ring parent moon:
\begin{equation}
a_{parent} = (G M_S)^{-1} \left( \frac{J_A + J_B + J_{MIMAS}}{M_A + M_B + M_{MIMAS}} \right) ^2
\label{eq-aparent}
\end{equation}
Assuming a uniform density $\Sigma_{ring}$ with radial range $R_0 < R < R_1$, the ring mass and angular momentum are:
\begin{eqnarray}
M_{ring} &=& \pi \Sigma (R_1^2 - R_0^2) \\
J_{ring} &=& \frac{4\pi}{5} (G M_S)^{1/2} \Sigma (R_1^{5/2} - R_0^{5/2})
\end{eqnarray}
For Saturn's $A$ and $B$ rings, the analysis of \citet{robbins10} and \citet{hedman16} sets limits 
on the ring surface densities with $\Sigma_A=420-520\;{\rm kg/m^2}$ and
$\Sigma_B=400-4800\;{\rm kg/m^2}$ though earlier estimates by \citet{esposito83} give $\Sigma_A=500\;{\rm kg/m^2}$ and 
$\Sigma_B\approx 1000\;{\rm kg/m^2}$. 
With the A and B ring radial ranges
of $122000-137000$ km and $92000-118000$ one can compute a range of ring masses and angular momenta using the equations above.
Using Mimas's mass and orbital radius of $M_{MIMAS}=3.75\times 10^{19}\;{\rm kg}$ and $a_{MIMAS}=185400\;{\rm km}$ once can
find the orbital angular momentum of Mimas and
so determine a mass and orbital radius for a ring parent moon from equation \ref{eq-aparent}.
The result is $M_{parent}=1.3-3.4 M_{MIMAS}$ with $a_{parent} = 128000-167000\;{\rm km}$ with the lower mass estimates corresponding to the
larger orbital radii.
Note that the mass of Enceladus is $M_{ENCELADUS}=2.9 M_{MIMAS}$ 
so the proposed ring parent moon is consistent with the properties of the innermost moons.
This orbital radius straddles the Roche radius defining the edge of the rings so in principle an ice+rock moon could be located 
in this position.  This orbital radius is also
fortuitously very close to the current 4:2:1 mean motion resonant orbital radius 
with Enceladus and Dione located at $a\approx 150000\;{\rm km}$.   
Mean motion resonances between satellites are common in the solar system and multiple
resonances have a precedent in Jupiter's moons with Io, Europa and Ganymede \citep{goldreich65} arranged 
in the same 4:2:1 hierarchy as originally observed by Laplace.  A reasonable hypothesis is that a similar resonant 
arrangement occurred with a ring parent moon, Enceladus and Dione when the inner icy moons of Saturn formed. 
Such a system of icy moons may have formed by the viscous spreading of a primordial massive 
ring composed of ice and rock
in the early phases of the formation of Saturn and its mid-sized moons \citep{canup10,charnoz10}.
Perhaps the last of this primordial ring material ended up in a moon just beyond the Roche radius.
If this hypothetical ring parent moon was trapped in resonance in the past,
its outward tidal migration would be slowed down significantly 
and it could remain in this vulnerable position prior to a collision that would eventually recreate a ring in recent
times.

A final requirement for this scenario to work is a mechanism that leads to the 
nearly pure icy composition
of Saturn's rings.
The hypothetical parent moon is likely differentiated from the strong tidal heating
expected from close proximity to Saturn and resonant interactions with Enceladus and Dione.
If we imagine a parent moon that is twice the mass of Mimas with the same rock mass then its ice
fraction is 84\%, a plausible value ranking it between Mimas and Tethys.
One might expect a partial disruption to liberate the mantle to form ice rings while
leaving behind the rocky core that migrates outward within a newly forming Mimas 
or small moon that subsequently interacts with Mimas.

In this paper, we explore this collisional scenario using simulations of hypervelocity cometary impacts with a
hypothetical differentiated ring parent moon located near Saturn's Roche radius.  We first describe
methods based on a new collisional N-body rubble pile code using simulations with $10^7$ particles.  
We present results that show how a ring of nearly pure ice can be created along with a remnant rocky
moon straddling the Roche radius.  We argue that the subsequent evolution of this system can 
plausibly transform into Saturn's ring system and a recently formed Mimas over a few hundred million years.
We also discuss the implications of this scenario for the rest of Saturn's inner satellite system.

\section{Methods}

\subsection{Collisional N-body methods}

We model the moon as system of gravitating hard colliding spheres using a new collisional N-body code based 
upon a parallelized N-body treecode originally developed for galactic dynamics \citep{dubinski96}.  
The code has been modified to follow collisions between particles treated as hard spheres and has similarities to other codes used 
to model rubble piles \citep[e.g.,][]{richardson00}.  The new code uses a parallelization
strategy that permits the simulation of N-body systems with as many as $10^8$ particles 
on modern parallel supercomputers.  We describe the code in detail in Appendix A.

\subsection{Ring parent moon}

We model the ring parent moon as a rubble pile of $N$ frictionless hard spheres of equal radius.
Following similar methods applied to models of asteroids \citep[e.g.,][]{leinhardt00}, we set up systems of 
hard spheres within an initially spherical distribution with small random velocities.  
For sufficiently large numbers of particles, impacts with rubble pile 
simulations behave like a fluid described by a hard-sphere equation of state (EOS) and therefore can be used as a 
proxy for more computationally difficult hydrodynamics simulations using equations
of state for ice and rock.  We justify this approximation for our problem in more detail below.
We describe the methods for setting up equilibrium gravitating systems 
of colliding hard spheres based on a hard-sphere EOS in Appendix B.

\begin{figure}
\includegraphics[width=6.5in]{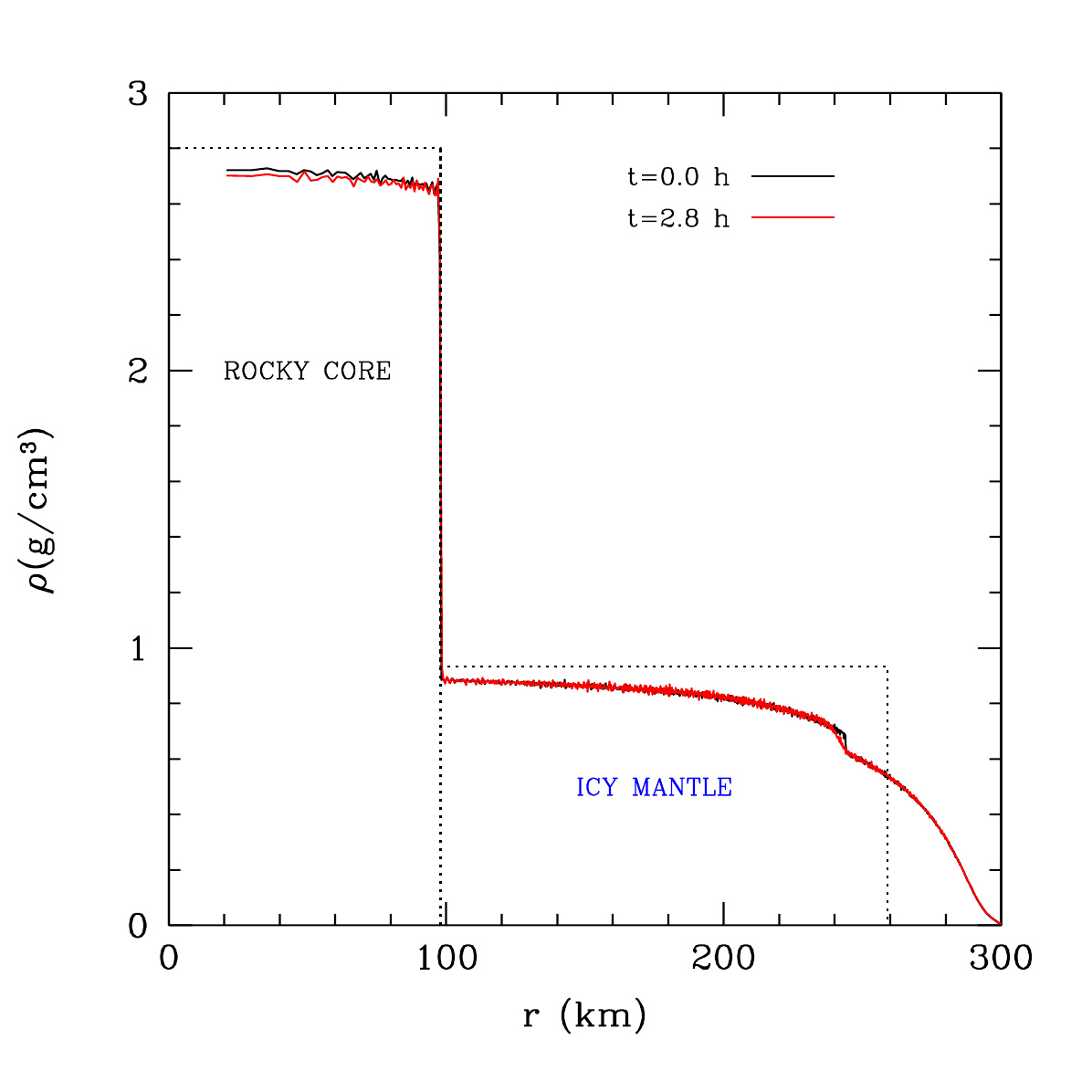}
\caption{The density profile of the ring parent moon model.  The moon has a rocky core with radius $R=100$ km and an icy mantle extending to $R=300$ km.
The dashed line describes the idealized density for a moon with a homogeneous rocky core and icy mantle.  The solid black line
describes the density profile for the hydrostatic equilibrium calculated from the hard-sphere equation of state with velocity dispersions of
$\sigma = 6\;{\rm m/s}$ in the core and $\sigma=11\;{\rm m/s}$ in the mantle.
The red line describes the density profile after
the model has been dynamically evolved for 2.8 hours showing that the model is almost in exact equilibrium.  
The discontinuity in density at $R=240\;{\rm km}$ is due to the phase change from
a hexagonally close packed crystal solid to a more random glass-like fluid phase.}
\label{fig-den}
\end{figure}

The ring parent moon model consists of 10M hard colliding spheres 
divided between a rocky core containing 
500K particles and a icy mantle containing 9.5M particles.
The particles within the rocky core have masses three times those of the icy mantle
so that mean densities of rock and ice are $\rho_{rock}=2.8\;{\rm g/cm^3}$
and $\rho_{ice}=0.935\;{\rm g/cm^3}$ - the density of ice at the ambient temperature of $T=70 K$ near Saturn.
The total mass of the moon is $M_{parent} = 2 M_{MIMAS}$ with the core mass intentionally
set to be the same value as the implied mass of rock within Mimas based on its mean density of $\rho=1.15\;{\rm g/cm^3}$.
The radius of the ring parent moon is $R=259\;{\rm km}$ leading to a mean density of $\rho=1.03\;{\rm g/cm^3}$ and
ice fraction of 84\%.  The density profile of our model is presented in Figure \ref{fig-den}.
The particle radii for this model are $R=1.05\;{\rm km}$.

\begin{figure*}
\includegraphics[width=6.5in]{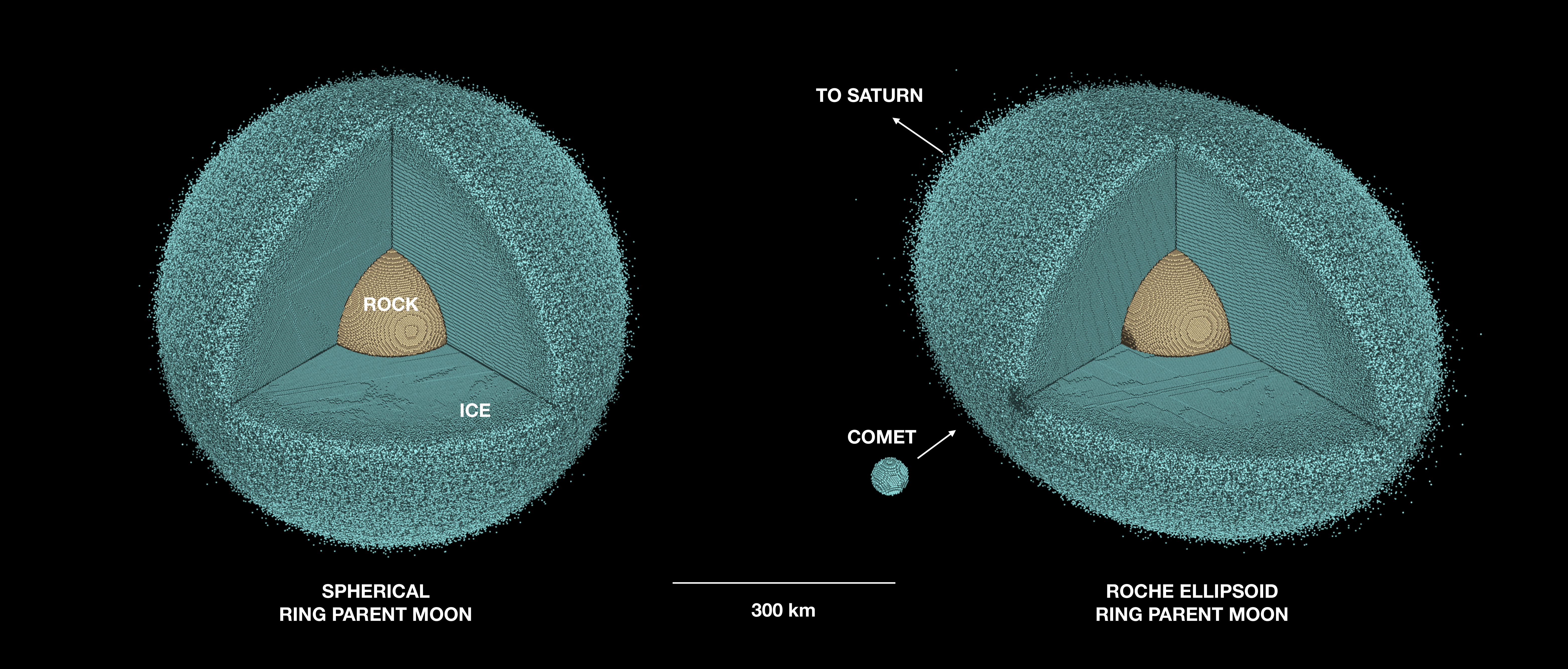}
\caption{A cutaway rendering of the spherical differentiated ring parent moon and the tidally adjusted Roche ellipsoid model with the comet impactor to scale.  
The moon model is composed of a rocky core containing $520K$ particles and icy mantle made of $9.5M$ particles.  All particles have a radius $R=1.05$ km.
The rock particles (orange) are $3\times$ the mass of the ice particles (blue).  The comet is made of 5000 ice particles.  The moon has a total mass
$M_{moon}=2 M_{MIMAS}$ while the comet has a mass $M_{c}=10^{-3} M_{MIMAS}$.}
\label{fig-moon}
\end{figure*}

The moon is placed in a circular orbit at $a=140000\;{\rm km}$ with
an orbital velocity $v_m=16.5\;{\rm km/s}$ and period $P=14.9$ hrs about Saturn.
For a moon this close to Saturn, one expects significant tidal flattening and the
moon should take the form of a Roche ellipsoid.  In principle, the exact shape of a Roche ellipsoid
can be computed for homogeneous bodies.  Since our model is differentiated, we find
the equilibrium configuration by setting the initially spherical model described above
in synchronous rotation on its orbit and letting it relax dynamically 
within Saturn's tidal field over one orbit.
The final Roche ellipsoid has an surface axis ratio 
a:b:c=1:0.69:0.65 and major axis length of $a=320\;{\rm km}$
while the inner rocky core remains nearly spherical (Figure \ref{fig-moon}).

\subsection{Comet}

We similarly construct a model comet as a spherical rubble pile made of 5000 ice particles.
The comet's mass is $M_c=3.75\times 10^{16}\;{\rm kg}$ with a radius $R=22\;{\rm km}$ comparable to the
size and mass of Comet Hale-Bopp \citep{weaver97}.  Below we show that this mass is sufficient to disrupt
a ring parent moon near the Roche zone.
The comet might come from the Oort cloud or the Centaur population.
For a zero-energy heliocentric orbit,
the excess speed at Saturn is $v_\infty = 14\;{\rm km/s}$. 
Centaur comets originating from Kuiper belt objects (KBOS) on Neptune crossing orbits 
have a median excess speed of approximately $v_\infty = 3\;{\rm km/s}$ as
determined from simulations (see Figure 1 of \citet{levison00}).
The comet speed before impact is
$v_{c}=(v_e^2 + v_\infty^2)^{1/2}\approx 23-27\;{\rm km}$ where $v_e$ is the escape velocity at the orbital radius $a$.
With the moon's orbital velocity of $v_m=16.5\;{\rm km/s}$, the relative velocity of impact $v_{rel}$ is 
in the range $7 < v_{rel} < 44\;{\rm km/s}$.  
In this paper, we consider three collisional trajectories corresponding
to the case of a rear-end, side-on and head-on collisions with relative velocities corresponding to comets on zero-energy orbits with 
$v_{rel}=11$, 27 and 44 km/s to explore a
range of impactor energies.  
In all three cases, the impacts are direct with no inclination to the surface and in the plane of the parent
moon's orbit.
In general, the comet could be coming in from any direction with a range of impact parameters and velocities.  The goal of this paper is to
demonstrate the plausibility of this mechanism for creating rings.  A more thorough study using a range of impact geometries and energies will
follow once the mass of the rings determined from the Cassini ring flybys is known.

\section{Collisional disruption}

An impacting comet deposits kinetic energy explosively at the surface of the target 
moon creating an inward propagating shockwave 
that leads to complete or partial disruption of the moon \citep[e.g.,][]{holsapple93}.
Collisional disruption is quantified by the parameter $Q_D^*$, 
the energy per unit mass of the target 
required to unbind half of the total mass \citep{benz99}.
For isolated pure ice targets, simulations show the scaling relation
$Q_D^*=0.05 (R/{\rm m})^{1.188}\;{\rm J/kg}$ where $R$ is the radius 
of the target in meters \citep{movshovitz15}.   
The typical mass of an impacting comet $M_{c}$ with 
relative velocity $v_{rel}$ that will disrupt the moon is found by equating the 
disruption energy to the kinetic energy
\begin{equation}
M_c = \frac{2 M_{moon} Q_D^*}{v_{rel}^2}.  
\end{equation}
A typical relative velocity is 27 km/s as discussed above requires a pure ice 
comet impactor mass, $M_c \approx 4\times 10^{16}\;{\rm kg}$ with
radius $R\approx 22\;{\rm km}$ to disrupt a ring parent moon of twice the mass of Mimas.  
We use a comet mass comparable to this value in our simulations.

The use of a collisional N-body code to follow this process needs to be justified since hypervelocity
disruptive impacts are complex and are usually treated with a hydrodynamics code using equations of state
for ice and rock that span behavior from the solid state through melting to vaporization \citep[e.g.,][]{benz99,kraus11}.
Collisional particle simulations do not
treat the detailed hydrodynamics and shock-heating 
induced phase transitions at the impact point
but we argue here that their application is a reasonable approximation since
the bulk of the energy goes into simply unbinding and shattering the moon.
One expects the shockwave from the impact to lead 
to the melting and vaporization of some 
of the target mass for impact velocities 
$> 10\;{\rm km/s}$ \citep{kraus11}.
For ice-on-ice hypervelocity collisions, the expected mass of 
melt plus vapor is $\sim 100$ times 
the mass of the impactor (see Fig. 6 in \citet{kraus11}).
The relatively small mass of the impactor in these simulations implies that around 
5\% of the mass of the moon is melted or vaporized. 
While collisional N-body simulations miss the details of the complex
behavior at the point of impact, most of the impact energy goes into 
disrupting the moon rather than melting and vaporizing it.
As the pressure wave from impact propagates away from the
impact point, the overpressure exceeds the local 
hydrostatic pressure and causes the moon 
to unbind.   A solid icy moon with a rocky core therefore
shatters into fragments 
and depending on the impact energy, the moon can completely or partially unbind.
For lower impact energies, a fraction of the mass 
can remain bound after the impact.  
The process of shattering is also not well-defined and depends on the structural
properties of the moon \citep{ballouz15} but in this rubble-pile approximation 
the moon simply breaks up into millions of equal mass fragments represented by
the particles.

These models use $10^{2-3}$ times as many particles as typically used in rubble pile
collisions and SPH simulations and so permit a small impactor to target mass ratio.   
In hypervelocity collisions with rubble piles, it is necessary to use small enough timesteps
to accurately distribute the impact energy throughout the target body and so conserve energy
through the disruption process.  Typically, the initial timestep needs 
to be $\delta t \approx R_p/v$ where $R_p$ is the radius of the particles and $v$ is the relative
velocity at impact.  For our largest simulations, $R_p=1.05$ km and $v < 44\;{\rm km/s}$ the initial $\delta t \sim 0.02 s$.  
Once the impact energy has distributed throughout the target body, the particle collision rate drops off as the density
decreases and the timestep can be increased to speed up the calculation.

To validate this approach, we carried
out a study of 20 isolated disruptive collisions of the putative ring parent moon using models composed of 1M particles with a range of impact 
energies varying both the mass ratio and relative velocity to explore the disruption process for comparison to hydrodynamics
calculations.   
During the impact and disruption, the particle collisions are elastic 
and non-dissipative with a coefficient of restitution $\epsilon=1$.
We are in effect using the rubble-pile code as a proxy to model a nearly incompressible dense fluid following a hard-sphere EOS
rather than an actual collection of boulders bouncing off of each other at low relative velocities in a non-destructive 
but dissipative manner.   For $\epsilon\approx 0.1-0.9$, as is often used in simulations of asteroid collisions, 
we determine in simulations that hypervelocity impacts dissipate energy at an artificially high rate and effectively quench disruption.
For elastic collisions, we find in the end that the total energy of the system is conserved to within 1\% for our choice of timestep.
In rubble-pile simulations,
the heating manifests itself as an increase in the velocity dispersion and decrease in particle 
number density as the hard-sphere ``gas" absorbs the energy.
After the disruption and dispersal of the mass along a ring, we use a coefficient of restitution $\epsilon < 1$ to follow the
dissipative evolution of the ring over 30 orbits when relative collision velocities are much smaller.

\begin{figure}
\includegraphics[width=6.5in]{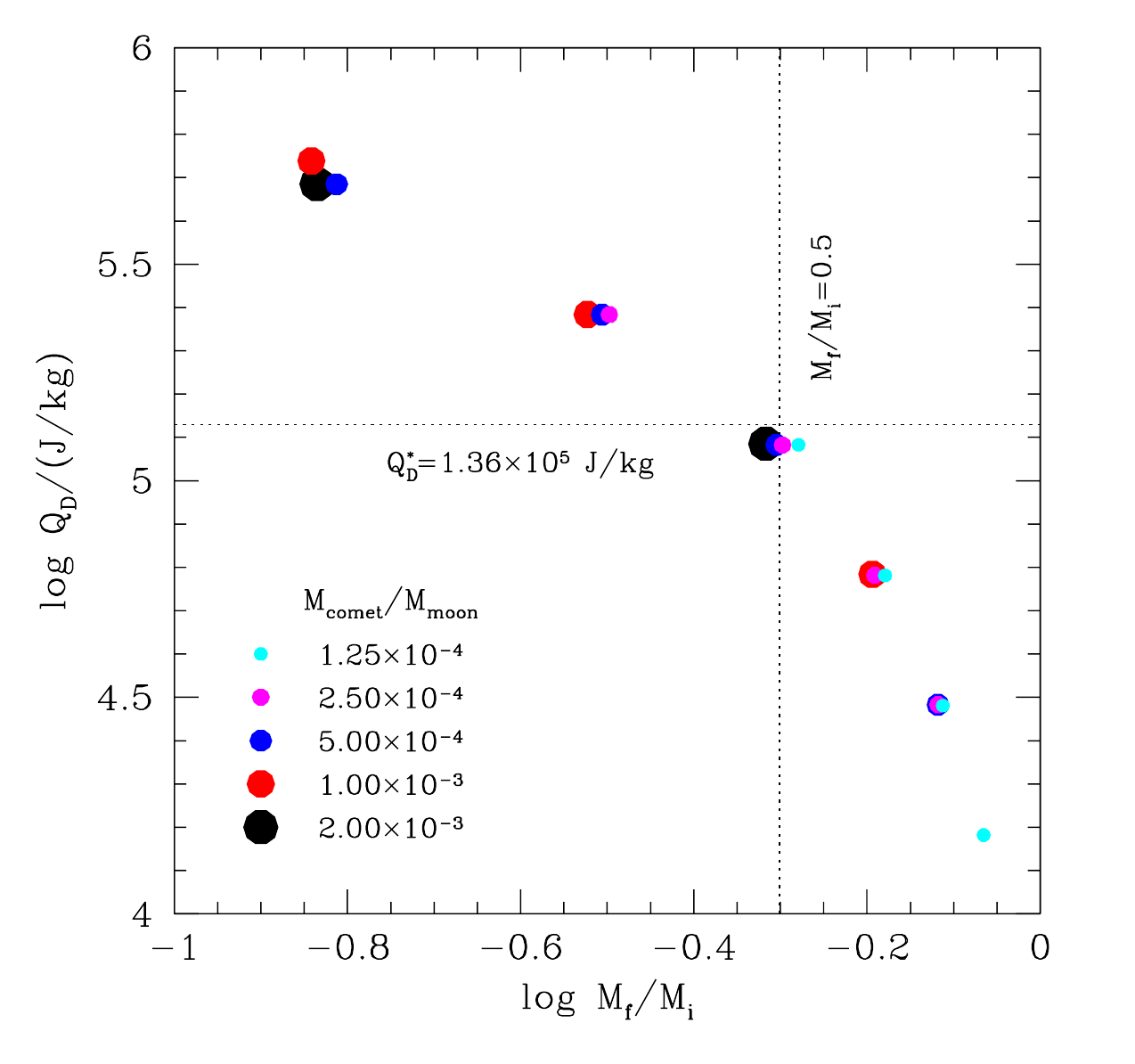}
\caption{The impact energy per unit mass versus the remnant bound mass for a range of hypervelocity impacts on the putative ring parent moon model.  Both the
mass and relative velocity of the impactor are varied to explore the behavior of the collisional disruption process.  
The groups of points at different impact energies reveal that
the controlling factor in collisional disruption is the energy per unit mass.  
The impactor to target mass ratio is less important as long as the ratio is
small.  
The plot also reveals that the critical disruption energy $Q_D^*$ for the ring parent moon agrees with the scaling relation of \citet{movshovitz15}.}
\label{fig-q}
\end{figure}

From the scaling relation of \citet{movshovitz15}, one expects $Q_D^*=1.36\times 10^5\;{\rm J/kg}$ for the ring parent moon model.
We used 5 relative velocities between 11-44 km/s in steps of $2^{1/2}\times$ 
and 4 comet masses ranging from $0.94-7.5 \times 10^{16}\;{\rm kg}$ in steps of $2\times$ corresponding to impact energies per unit mass ranging
from $Q_D=0.015-3.8\times 10^6\;{\rm J/kg}$.  In each case, we followed the collision well past the impact and partial 
unbinding of mass to the re-accretion of material into a bound body.  
We estimated the mass of the bound remnant using a friends-of-friends particle grouping method
similar to algorithms used for finding halos in cosmological simulations.  
Figure \ref{fig-q} plots the collision energy per unit mass $Q_D$ versus the remnant
bound mass for the different comet:moon mass ratios.   The plot shows a grouping of simulations 
near the predicted value of $Q_D^*=1.36\times 10^5\;{\rm J/kg}$ and
a remnant mass of 50\% of the initial mass.  The implication is that rubble pile simulations with hypervelocity 
impacts reproduce the results of hydrodynamic simulations of disruption.
More complex hydrodynamic simulations should be done eventually to validate these findings
but this set of experiments gives us confidence that the use
of rubble piles in this context is a reasonable approximation.

\section{Ring parent moon disruption in orbit about Saturn}

\begin{figure*}
\includegraphics[width=6.5in]{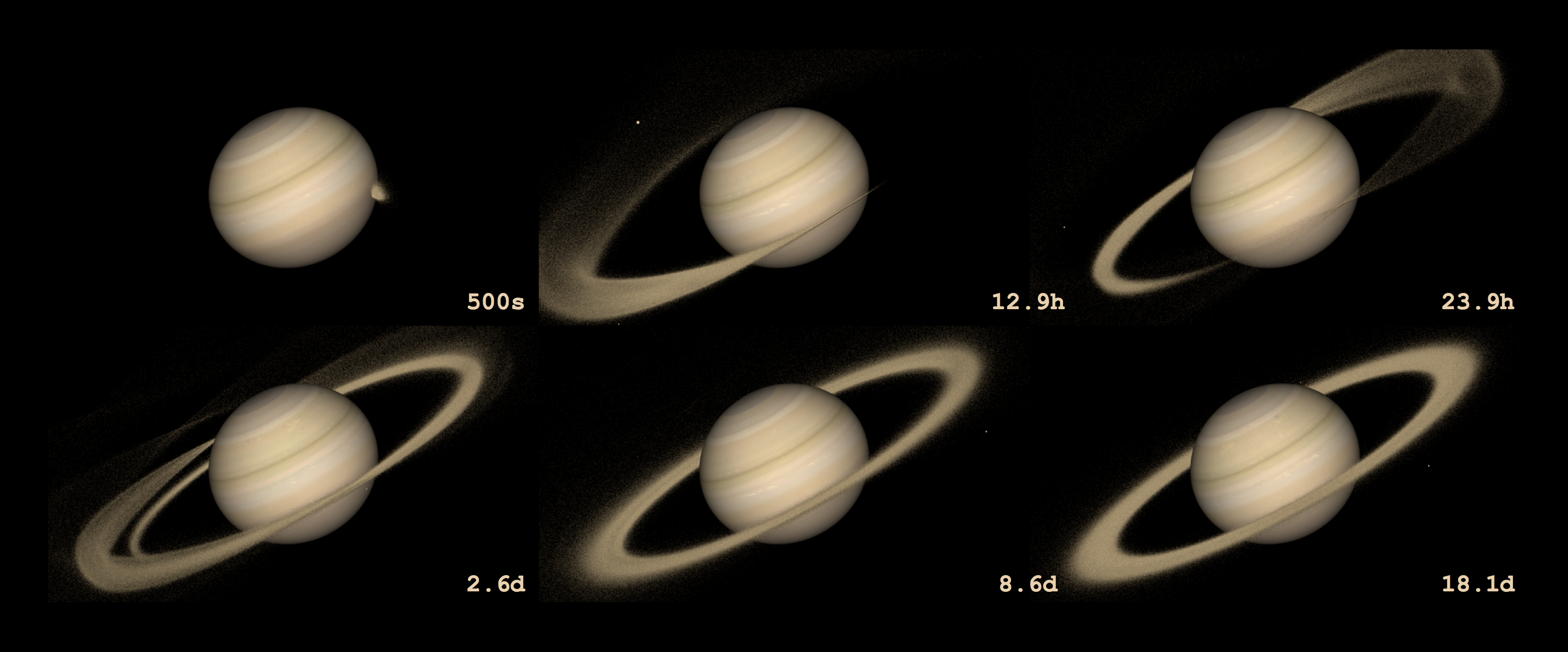}
\caption{The formation of a ring around Saturn from the disruption of a parent icy moon after a collision with a comet.  
The moon is initially on a circular orbit with radius $a=140000\;{\rm km}$ with an orbital period of 14.9h.  
The moon has a mass of $M=7.5\times 10^{19}\;{\rm kg}$ and is differentiated with an icy mantle containing 84\%
of the mass and a rocky core containing the remaining mass.
In this example, a comet of mass $M_{comet}=3.75\times 10^{16}\;{\rm kg}$ collides with the moon head-on 
within the orbital plane with relative velocity 
$v_{rel}=44\;{\rm km/s}$ and disrupts the moon leaving a remnant rocky core containing 8\% of the original mass.  
The debris spreads along the orbital radius with inelastic collisions between particles leading to a thin ring in 
Saturn's equatorial plane within a few weeks.  At the end of the simulation, the ring is composed 
of 91\% ice with a radial width of approximately 10,000 km. See accompanying animation at this link: https://youtu.be/UtVnftTd1tA}
\label{fig-saturnseq}
\end{figure*}

We now proceed to simulate the disruption of the ring parent moon modeled as a Roche ellipsoid in orbit about Saturn.  
Saturn's gravitational field is modeled as a background potential within the N-body code.  The potential is
represented as a spherical harmonic expansion using 
Saturn's mass and zonal harmonics from Table 3 of \citet{jacobson06} to quadrupole order.
We examine three representative collisional scenarios where the comet impacts the 
moon within the orbital plane from the rear-end (trailing face), side-on and 
head-on (leading face) directions corresponding to different relative velocities and impact energies (Table \ref{table-sim}).
\begin{table*}
\centering
\begin{tabular}{rrrrrr}
\hline
Collision & $v_{rel}$ (km/s) & $Q_D$ ($10^5$ J/kg) & $M_{remnant,i}/M_{Mimas}$ & Ring ice fraction \\
\hline
rear-end &  11.2 & 0.3 & 1.02 & 1.0& \\
side-on  &  22.4 & 1.2 & 0.74 & 1.0& \\
head-on  &  44.2 & 4.9 & 0.16 & 0.91& \\
\hline
\end{tabular}
\caption{The mass of the comet is $m_{comet} = 3.75\times 10^{16}\;{\rm kg}=5\times 10^{-4} m_{moon}$ and it is 
100\% ice.
The remnant mass is measured at the end of the simulation. The rocky core remains intact for the rear-end 
and side-on collisions while about half of the rocky core in the head-on collision ends up in the ring.  One expects the
remnant to accrete some of the mass of the nascent ring as it spreads.   The optimal impact energy
for creating Mimas with the right mass is somewhere between $1.2-4.9\times 10^5\;{\rm J/kg}$.  The rear-end collision produces a
remnant that is greater than the mass of Mimas and can be excluded as a plausible scenario.}
\label{table-sim}
\end{table*}
Figure \ref{fig-saturnseq} 
and the accompanying animation
present the result of the head-on simulation and illustrate the disruption after impact
and the ring formation process.  

\begin{figure*}
\includegraphics[width=6.5in]{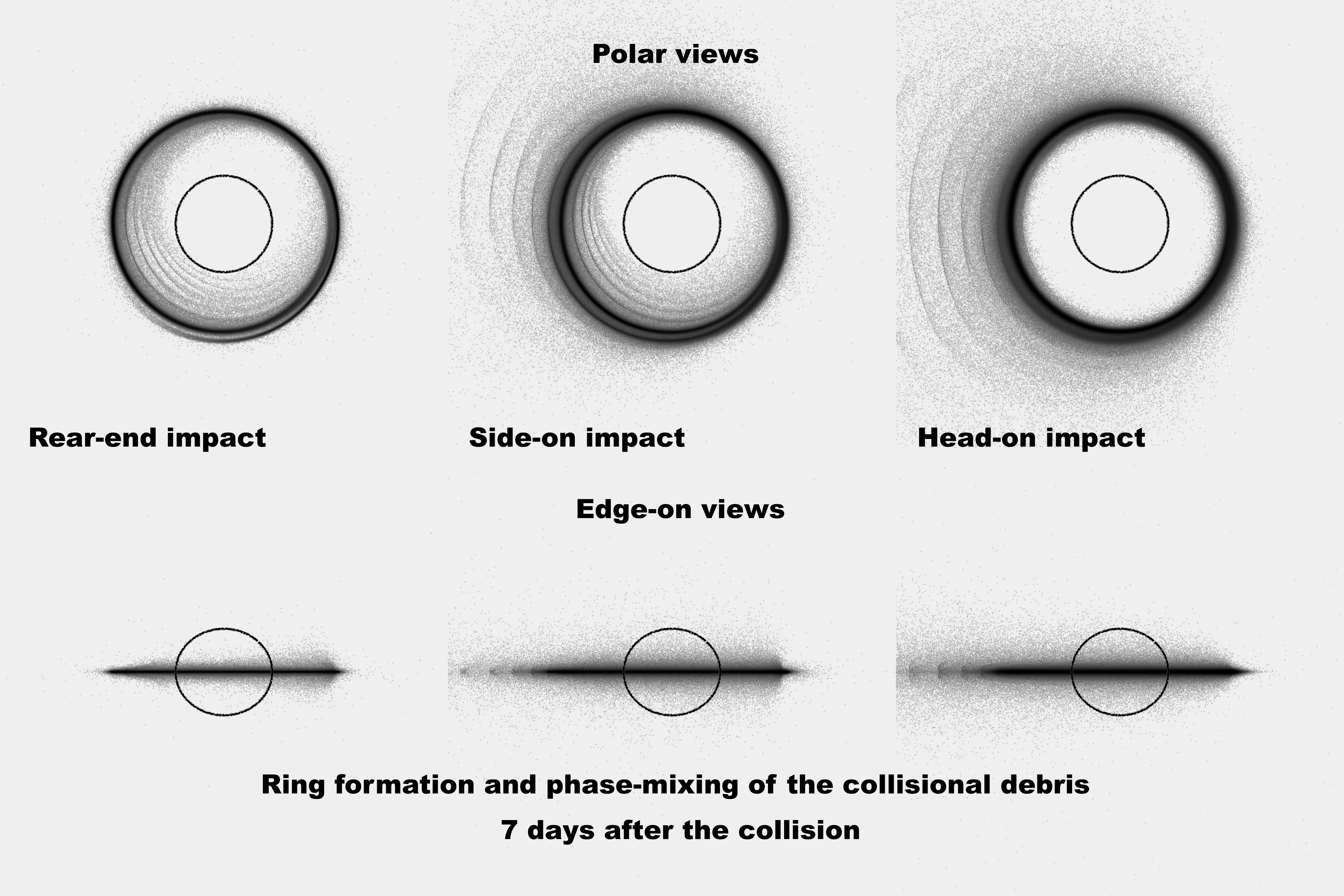}
\caption{The distribution of collisional debris and the formation of a ring in the three scenarios where the comet impactor collides with
the ring parent moon from the rear, the side and head on.   The comet orbit is coplanar with the target moon.  The polar and edge-on views
of the debris are shown 7 days or approximately 11 orbits after the impact.  The central circle or ellipse represents Saturn to scale.
In the rear-end collision, the impact ejecta trails the orbit and falls into Saturn and phase mixes.   
Particles pile up on orbital turning points leading to grooves in the distribution.  
For the side-on collision, debris trails and leads the orbit leading to grooves inside and
outside of the orbit.  In the head-on collision, the debris leads the orbit and the grooves are outside the orbit.
Over time, the groove features are erased as the particles on eccentric orbits circularize after successive collisions with the forming ring. 
The simulations end after about 20 days or 30 orbits and a broad featureless ring develops by this time.  See the animation of this process at this link: https://youtu.be/t8GtvzyD0xw }
\label{fig-rings-phasemix}
\end{figure*}

\begin{figure*}
\includegraphics[width=6.5in]{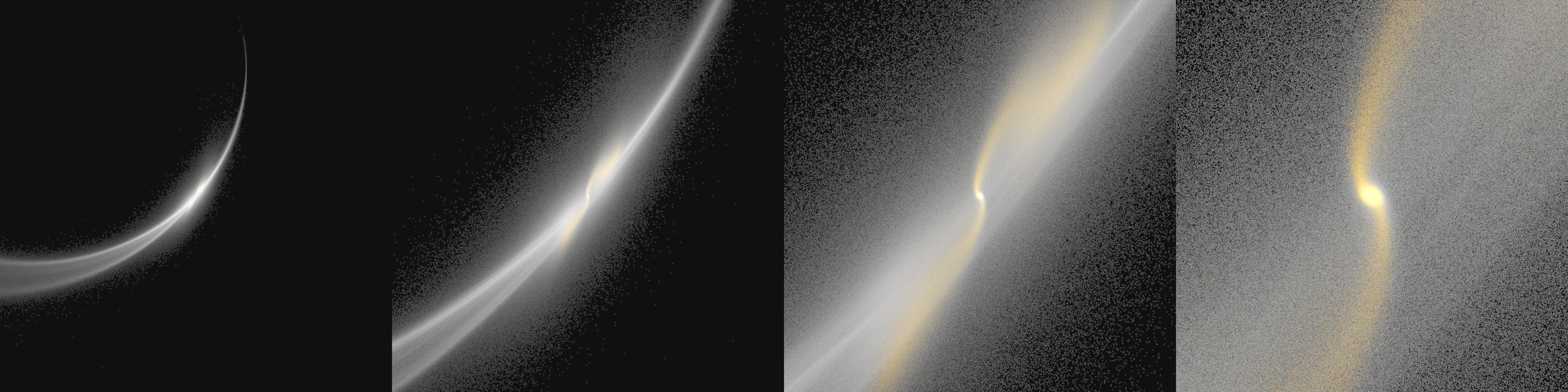}
\caption{Zooming in on the collisional remnant in the head-on collision
simulation one orbit after the impact.  
Ice particles are rendered in white and
rock in red.   The width of the images from left to right are 288000, 72000,
18000, and 4500 km.   The red central spheroid of the remnant is
approximately 300 km across.
This simulation is the most
energetic of the three and almost disrupts the moon entirely. 
All of the icy mantle of the ring parent moon is ejected
while leaving behind the rocky core.  Rock debris leaks away from the Roche
lobes of the nearly disrupted moon and mixes with the forming ring
leading to an ice fraction of 91\% consistent with the inferred composition
of the ring.}
\label{fig-remnant}
\end{figure*}

In all cases, the comet is obliterated on impact and deposits its energy in a point-like explosion 
on the surface of the target moon.  The impact ejecta escapes the moon and spreads out along the orbit (Figure \ref{fig-rings-phasemix}).
In the head-on collision, the moon is almost completely disrupted
and the impact ejecta leads the orbit ending up on eccentric trajectories with apogees larger than the initial orbital
radius.    After one orbit, only 8\% of the mass remains in a bound object composed of about half of the original rocky core
with all of the icy mantle liberated to form a ring (Fig. \ref{fig-remnant}).

In the side-on and rear-end collisions, a significant fraction of the icy mantle becomes unbound but the rocky core remains
intact with a depleted icy mantle in remnants which contain a mass of 37\% and 51\% of the original moon respectively (0.74 and 1.02 $\times$ the mass of Mimas).
The remnant bound mass depends on the 
impact energy as shown in collisional disruption studies \citep{movshovitz15} though it seems for the lowest energy
impact corresponding to the rear-end collision significantly more mass is lost compared with isolated targets.
In the case of a moon near the Roche radius in orbit around Saturn, 
more mass can become unbound for a given impact energy than
after the collision because of the small size of the Hill sphere.   
In these simulations, the Hill radius of the parent moon
is only $700\;{\rm km}$ -- just a few times the moon's radius.
Debris that would normally remained bound to a disrupted moon and re-accrete can instead
escape when it moves beyond the Hill radius and mix with the forming ring system. 
Lower energy impacts can therefore still be effective at unbinding a large fraction of the mass and forming a ring.

After the moon disrupts, the debris settles along the orbit to form an 
initial squashed toroidal distribution of particles with a radial width $\sim 10000$ km (full-width half maximum of the surface density) 
and a RMS scale-height of $\sim 1000$ km.  The evolution of this system is followed for 30 orbits.
Once the debris disperses along the orbit, we allow the collisions to be inelastic using
coefficient of restitution $\epsilon=0.9$ with a slight modification to permit elastic collisions
with $\epsilon=1.0$ for relative velocities with $v<50\;{\rm m/s}$ to prevent runaway clumping
of particles (see Appendix A).  
This value of $\epsilon$ is in accord with laboratory measurements of low velocity impacts of ice spheres
\citep{higa96}.  
To save computation time, the remnant moon is extracted after one orbit and replaced with a single sink particle 
moving along the same trajectory.  Particles that collide with this sink particle are removed from the simulation
and their mass and momentum are absorbed.  
The initial velocity dispersion is $\sigma \sim 0.5\;{\rm km/s}$ so collisions of particles in the forming
rings are energetic enough to initiate a collisional cascade \citep{dohnanyi69} that will break up the particles 
into smaller pieces as they dissipate energy inelastically.  Collisions of ice projectiles in experiments 
with relative velocities $>10\;{\rm  m/s}$ are expected to shatter \citep{hartmann78}.
The number of particles should grow and lead to an increase in the ring optical depth and collision rate until most of
the initial random kinetic energy from the disruption is dissipated in inelastic collisions.

It is not possible to simulate a collisional cascade with the existing rubble pile code so instead we
use an {\em ad hoc} method to approximate the expected growing optical depth $\tau = N \pi r_p^2$ where $N$ is the 
particle surface number density and $r_p$ is the simulation particle radius.
We increase the optical depth of the forming rings artificially by 
growing the cross sectional area (increasing the radius) of the particles
in two stages after the debris settles into a ring.   
In this way, the simulation particles are transformed into
scaled-particles to increase the ring optical depth and collision rate using a
fixed number of particles \citep{rein10}.  
As the ring evolves and spreads radially, the optical depth drops.  When $\tau \approx 0.1$,
the particle cross sectional areas are enlarged $10\times$
to bring the optical depth back to $\tau \approx 1$.
This greater collision rate emulates the effect of a collisional cascade and speeds up the
dissipation of energy and relaxation of the ring.
The dissipation of energy from
inelastic collisions between the particles cause the ring to spread radially 
and become vertically thinner \citep{brahic77}.
The ring particle 
orbits also undergo orbital phase mixing spreading azimuthally to 
form a ring with an approximate Gaussian 
surface density profile (Figure \ref{fig-rings-phasemix}).
Given the idealized nature of these calculations, the appropriate value for $\epsilon$ is uncertain and the chosen rate 
of growth of the ring optical depth to emulate the effect of a collisional cascade is also not well-defined.  
At the end of these simulations, the velocity
dispersions are still relatively high with $\sigma \sim 50\;{\rm m/s}$ (see below) 
much larger than the current value of $\sigma \sim 0.01;{\rm m/s}$ \citep{goldreichtremaine78}.  
Despite the uncertainties, the rate of dissipation
in the nascent ring in this early stage may be over-estimated but one
expects the structural evolution to proceed in qualitatively 
the same way since angular momentum is conserved.  

\begin{figure*}
\includegraphics[width=6.5in]{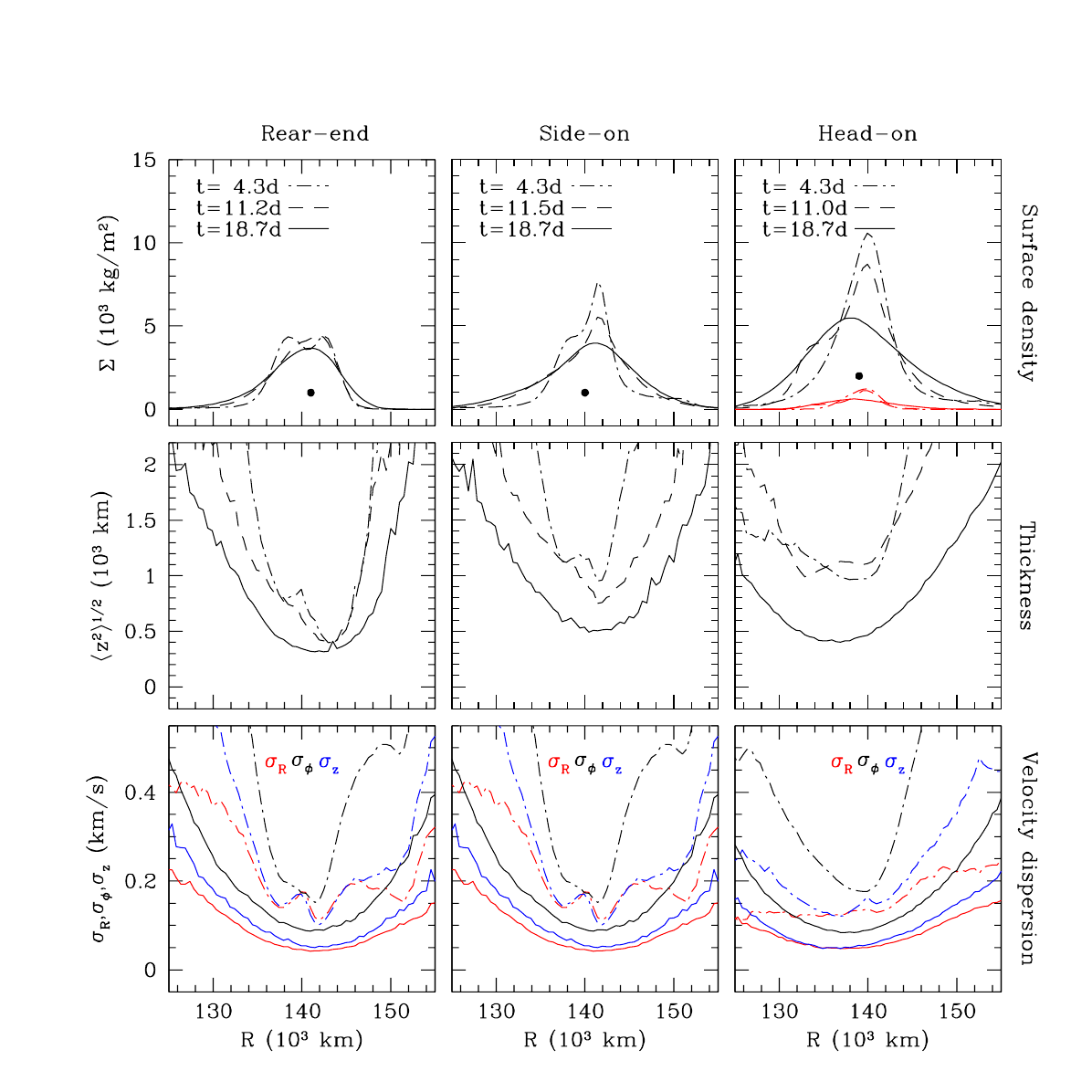}
\caption{Structure and kinematics of the ring for the three collision cases: rear-end, side-on and head-on
at three times during the formation and evolution of the ring.
The structure is measured from the azimuthally averaged surface density and the thickness as measured by the RMS of 
the particles vertical position with respect to the equatorial plane.  The kinematics are described by the
azimuthally averaged velocity ellipsoid in different radial bins.  Within a few weeks, the ring surface density
has a Gaussian shape with maximum centered on the initial orbital radius and a full width half maximum of 
about 10000 km.  Collisional dissipation leads to the circularization of the ring particle orbits and the decline
in velocity dispersion to $0.05-0.1\;{\rm km/s}$.
In each case, the disruption is not complete and a remnant remains in orbit embedded within the debris
- the radial position of the remnant's semi-major axis is shown by the black dot.  For the rear-end and side-on
collisions, the remnant contains nearly half of the original rocky core and the ring is pure ice.  For the head-on
collision, the disruption is nearly complete but a remnant core containing half the rock remains with the
debris (shown in red in the surface density plot) mixing with the ring.}
\label{fig-ringprofiles}
\end{figure*}

Figure \ref{fig-ringprofiles} shows the surface density profile, scale height and velocity dispersion of the ring at three times for the 
head-on collision and reveals the rapid transformation of the spreading debris cloud from a point-like 
collisional disruption into a ring system.  
The final state of these simulations can be thought of the initial conditions for the subsequent 
long term phase of viscous spreading of the rings into their current state as described by \citep{salmon10}.
The three simulations span a range of impact energies and each form a ring system with a high ice fraction (see Table 2).
For lower impact energies, a significant bound mass made predominantly of the original rocky core can survive
while the debris is composed of the icy mantle.  This is a simple way to explain the 
icy purity of the rings though it requires an initial moon that is differentiated with a relatively high ice fraction of 84\% --
albeit in the expected range for Saturn's moons.   
Over the long term, the remnant should accrete a significant fraction of the nascent ring as it spreads viscously.
The least energetic rear-end collision produces a remnant that is greater than the mass of Mimas and so can be excluded as a possibility.
The other two models have remnants with less mass than Mimas that will grow in size by accretion of the spreading debris.
A more thorough study involving moons with different ice
fractions and cometary impact energies can set stronger constraints on the range of plausible scenarios once the mass of the rings
are determined from the Cassini ring flyby data.  
The 3 simulations presented here demonstrate that it should be relatively easy to find simulation parameters
that preserve a rocky core while liberating an icy mantle.  

The remnant acts as a new seed for the re-accretion of debris to form a new moon which one expects to clear 
out the debris within its orbital radius within a timescale as short as $10^3$ years \citep{crida12}.
The long term evolution of a collision remnant embedded within a newly formed ring should be followed to find scenarios that
will produce a Mimas of the right size while some of the debris moves into the Roche zone.
One expects the newly formed Mimas to
experience resonant interactions with the particles in the spreading ring.  The resulting torques lead to outward
migration with the orbital radius $a$ reaching its current value with a timescale:
\begin{equation}
\tau_{migration} = \left ( \frac{1}{a}\frac{da}{dt} \right )^{-1} = 0.60 \frac{M_{Saturn}^2}{a^2 \Sigma \Omega M_{Mimas}}\left |\frac{a-r}{a}\right |^3
\end{equation}
where $\Sigma$ is the surface density of the ring, $\Omega$ is the orbital frequency and $r$ is the radius of the 
outer edge of the ring \citep{goldreichtremaine82}.
From these simulations,  the initial central surface density after settling down into a thin ring 
is $\Sigma \approx 5000\;{\rm kg/m^2}$.   With the ring's outer edge at $r=140000;{\rm km}$ and the current
radius of Mimas' orbit at $a=185400;{\rm km}$ lead to a migration
timescale of $\tau_{migration} = 120\;{\rm Myrs}$.  In the region between the rings and Mimas, torques from ring interactions are greater than the torques from tidal
interactions with Saturn \citep{crida12}.  Since the mass of the ring and Mimas are comparable, the 
outward migration of Mimas will result in a back reaction from the ring which will cause it to spread inward more rapidly 
than expected from collisional viscous effects alone and its mean surface density will decline as the ring spreads
to the current estimated value of $\Sigma = 1000\;{\rm kg/m^2}$ for the B ring.   
This sets a more conservative timescale of a few times $\tau_{migration}$ because of the inverse dependence on $\Sigma$.
The recent value from \citet{lainey17} for Saturn's tidal dissipation factor $k_{2,S}/Q_S \approx 1.6\times 10^{-4}$ predict a tidal migration time of
a Mimas from $a=140000$ km to its current orbital radius in only 450 Myr.   
More detailed dynamical calculations of this system are required
but the timescale for moving Mimas out from the edge of the rings to its current position in
a few hundred million years consistent with the assumption of a recent formation time for the rings.

\section{Discussion}

The mechanism responsible for the creation of Saturn's rings is intertwined with the origin and subsequent dynamical evolution 
of Saturn's icy mid-sized moons.  In this section, we discuss some of the implications for the Saturn system
from the proposed collisional disruption mechanism.

\subsection{The age of Mimas}

The scenario proposed in this paper suggests a coeval origin of Saturn's rings and Mimas within the past few hundred million years.
This is contrary to the conventional view that Mimas formed primordially with Saturn and the other inner mid-sized icy moons more 
than 4 billion years ago.
The heavily cratered surface of Mimas is consistent with a very old age based on measurements of the size distribution of craters 
if they originate
from objects on heliocentric orbits \citep{zahnle03}.
However, measurements of the libration of Mimas by the Cassini images \citep{tajeddine14} are at odds with a primordial origin.
If Mimas formed primordially, one expects it to be either a homogeneous or differentiated body in hydrostatic equilibrium but
direct measurement of libration show anomalies inconsistent with these states.  
Two explanations that can explain the anomaly are 1) an internal ocean or 2)
a non-hydrostatic ellipsoidal rocky core.  We have shown that an icy ring can form from the partial disruption of a differentiated 
ring parent moon leaving
behind a rocky core that subsequently re-accretes to become Mimas while being pushed out to its current position through 
a combination of resonant
interactions with the ring and tidal dissipation.  It seems plausible that the post-impact disturbed remnant rocky 
core would not have time to relax 
to hydrostatic equilibrium and so support this hypothesis as the origin for the anomalous libration.   
Recent models of the thermal, structural and orbital
evolution of Mimas are consistent with a late, layered formation scenario similar to the one proposed here \citep{neveu17}.

\subsection{Mean motion resonances and the heating of Enceladus}

The ring forming scenario has a further implication that may help in understanding another puzzle in Saturn's system of moons:
the heat source of Enceladus endogenic activity.  \citet{spencer06} have measured the current heat output from Enceladus as $5.8\pm1.9\;{\rm GW}$ which is
much larger than the heat input expected from either radiogenic sources in the rock or tidal heating from the current MMR with Dione.  
If a ring parent moon is in MMR
with Dione and Enceladus prior to disruption, one expects the transfer of angular momentum outward from tidal friction
to excite a larger eccentricity in the orbit of Enceladus which in turn increases the rate of tidal heating \citep{squyres83,peale99}.
It has been argued before that
the outward transfer of angular momentum from a possible past orbital resonance with Janus may have pumped up the
eccentricity of Enceladus thus providing a heat source \citep{lissauer84}.   The mass of Janus proved to be too small
for this heating mechanism to work but the ring parent moon proposed here may have enough mass to increase
the tidal heating rate to values significantly above the current rate.

To illustrate this, we assume that the ring parent moon, Enceladus and Dione are in MMR prior 
to the comet collision with orbital frequencies in the ratio of 4:2:1.
While in resonance, tidal friction will cause the moons to migrate outwards in tandem while maintaining the ratio 
of orbital frequencies and semi-major axes.  
For 3 moons in MMR, the time for the innermost moon to migrate from semi-major axis $a_i$ to $a_f$ is given by:
\begin{equation}
t = t_{migrate} \left [ 1 - \left ( \frac{a_i}{a_f} \right )^{13/2} \right ]
\end{equation}
where the time constant is:
\begin{equation}
t_{migrate} = \left [ \frac{39}{2} \frac{m_1}{M_S} \left ( \frac{R_S}{a_1} \right )^5 \frac{k_{2,S}}{Q_S} n_1 \eta\right ]^{-1},
\end{equation}
with
\begin{equation}
\eta = \frac{1 + (m_2/m_1)^2 (a_2/a_1)^{-6} + (m_3/m_1)^2 (a_3/a_1)^{-6}}{1 + (m_2/m_1)(a_2/a_1)^{1/2} + (m_3/m_1)(a_3/a_1)^{1/2}}
\end{equation}
and $m_1$, $m_2$, $m_3$ are the masses of the moons, $a_1$, $a_2$, and $a_3$ are semi-major axes of the moons' orbits, 
$n_1$ is the orbital frequency of the innermost moon, $R_S$ and
$M_S$ are the radius and mass of Saturn and $k_{2,S}$ and $Q_S$ are Saturn's Love number and dissipation 
function \citep[e.g.,][]{murray00}.   
The conventional value for Saturn's dissipation term is $k_{2,S}/Q_S=2\times 10^{-5}$ calculated under the assumption that Mimas
migrates from the synchronous radius to its current radius over 4.6 Gyr (the age of the solar system) by tidal evolution \citep{goldreich66}.
With this value, the time for the ring parent moon considered here in MMR to migrate from an orbital radius of $a=130000$ km to 150000
km is 4.5 Gyr showing that it is possible to trap such a moon near the Roche radius for the age of the solar system.
For $k_{2,S}/Q_S=1.6\times 10^{-4}$ proposed by \citet{lainey17}, this timescale is only 0.6 Gyr. 
However, Lainey's value for Saturn's dissipation term
implies a recent origin for most of Saturn's icy moons, 
since all moons out to Dione's orbital radius would migrate to their current locations in less than 2 Gyr \citep{cuk16}.

Assuming nearly circular orbits and conservation of energy and angular momentum, 
one can also calculate the net heat flow into the moons from orbital binding
energy as:
\begin{equation}
H = \sum_i n_i T_i + \frac{2\sum_i T_i \times \sum_i E_i}{\sum_i L_i}
\label{eq-H}
\end{equation}
where $n_i$ are the satellite orbital frequencies, $T_i$ are the torques on the satellites due to tidal evolution, $E_i$ and $L_i$ are the satellite orbital
binding energies and angular momenta \citep{lissauer84,meyer07}.  
The torque from tidal evolution is given by:
\begin{equation}
T = \frac{3}{2} \frac{G M_m^2 R_S^5}{a^6} \frac{k_{2S}}{Q_S}
\label{eq-torque}
\end{equation}
where $M_m$ is the mass of the moon, $a$ is the orbital radius and $R_S$ is the radius of Saturn.
A system of satellites in MMR will excite orbital eccentricities in the moons with equilibrium values that
balance the heat gain from the tidal evolution of the orbits and the heat loss from internal dissipation 
in the moons caused by the varying tidal field on the
eccentric orbit.  
The dissipation rate for a synchronously rotating moon is:
\begin{equation}
\frac{dE}{dt}=\frac{21}{2}\frac{k_{2,m}}{Q_m} \frac{G M_S^2 n R_m^5}{a^6}e^2
\label{eq-peale}
\end{equation}
where properties of the moon are denoted with the subscript $m$ and $e$ is the orbital eccentricity assumed to be small \citep{peale99}.
(For the more general formulae, see \citet{wisdom08} - for $e=0.1$ the dissipation rate is about 20\% larger than that given by Eq.~\ref{eq-peale}).
The satellite Love number can be estimated by:
\begin{equation}
k_{2,m} = \frac{3/2}{1 + 19\mu/(2\rho g R)}
\label{eq-love}
\end{equation}
with rigidity $\mu = 4\times 10^9\;{\rm N m^{-2}}$ and $\rho$ is the density, $g$ is the surface acceleration and $R$ is the radius with the
satellite dissipation term in the range $Q_m=20-100$ \citep{meyer07}. 

For Enceladus and Dione in a 2:1 resonance, the above analysis was used to estimate a heating rate of 1.1 GW \citep{meyer07} which is much less than the
measured heat output of Enceladus \citep{spencer06}.   If a ring parent moon existed in the past, this heating rate would be significantly larger.
Let us consider a parent moon that is twice the mass of Mimas $M=7.5\times 10^{19}\;{\rm kg}$ used in our simulation that begins 
at an orbital radius of $a=135000$ km just beyond the Roche radius with Enceladus and Dione in MMR with orbital frequencies in the ratio 4:2:1.
Using the above equations with $Q_m=20$ for the moons, one finds a heat flow from equation \ref{eq-H} into the 3 moons of $H=107\;{\rm GW}$ (and $H \approx 1000$ GW for Lainey's $Q_S$!) .  
This is 2
orders of magnitude larger than the current tidal heating rate estimated above.
Prior to the ring parent moon's destruction, this higher heating rate on Enceladus implies
that its icy fraction may have been completely molten creating a smooth crater-free surface \citep{smith82}.
After the disruption of the parent moon and the formation of 
the ring, the eccentricity of
Enceladus would decay to its current value with a time constant of order $10^8$ years \citep{meyer07} 
with the heating rate falling off and Enceladus refreezing.
It has been argued that a previous molten state driven by a more eccentric orbit in the past
is needed to explain the observed heat output
under the more modest tidal heating rate existing today \citep{ojakangas86}.

To further understand this scenario, we carried out orbital integrations of an idealized co-planar system containing 
a ring parent moon, Enceladus and Dione in MMR to illustrate both the long term stability of this system 
and heating rates.  We modified the Mercury N-body code \citep{chambers99} 
to include velocity-dependent fictitious forces to mimic tidal evolution for the moons \citep[e.g.,][]{lee02}.  Torques from
tidal evolution and internal dissipation are included allowing an exploration of the evolution of the semi-major axis, eccentricity and tidal heating of the
satellites.  
For Saturn, we use both the conventional value of $k_{2,S}/Q_S=2\times 10^{-5}$ for the
tidal dissipation term as well as the recent \citet{lainey17} value of $k_{2,S}/Q_S=1.6\times 10^{-4}$.   
We compute the satellite Love numbers from equation \ref{eq-love} with different values of tidal dissipation $Q_m$ for the moons.
Since tidal migration is very slow compared to the orbital period, we speed up the calculations by increasing
the $Q$ factors by $100\times$ \citep[e.g.,][]{meyer08} and rescaling the timescale in the plots \citep{meyer08} 
(Note: calculations with a speed up of $1000\times$ also gave the same result).  
We begin the integration with the parent moon on a nominally circular co-planar orbit with $a=135000$ km and Enceladus and Dione with
orbital radii slightly larger than their $2:1$ and $4:1$ resonant positions with respect to this initial radius.

\begin{figure}
\includegraphics[width=6.5in]{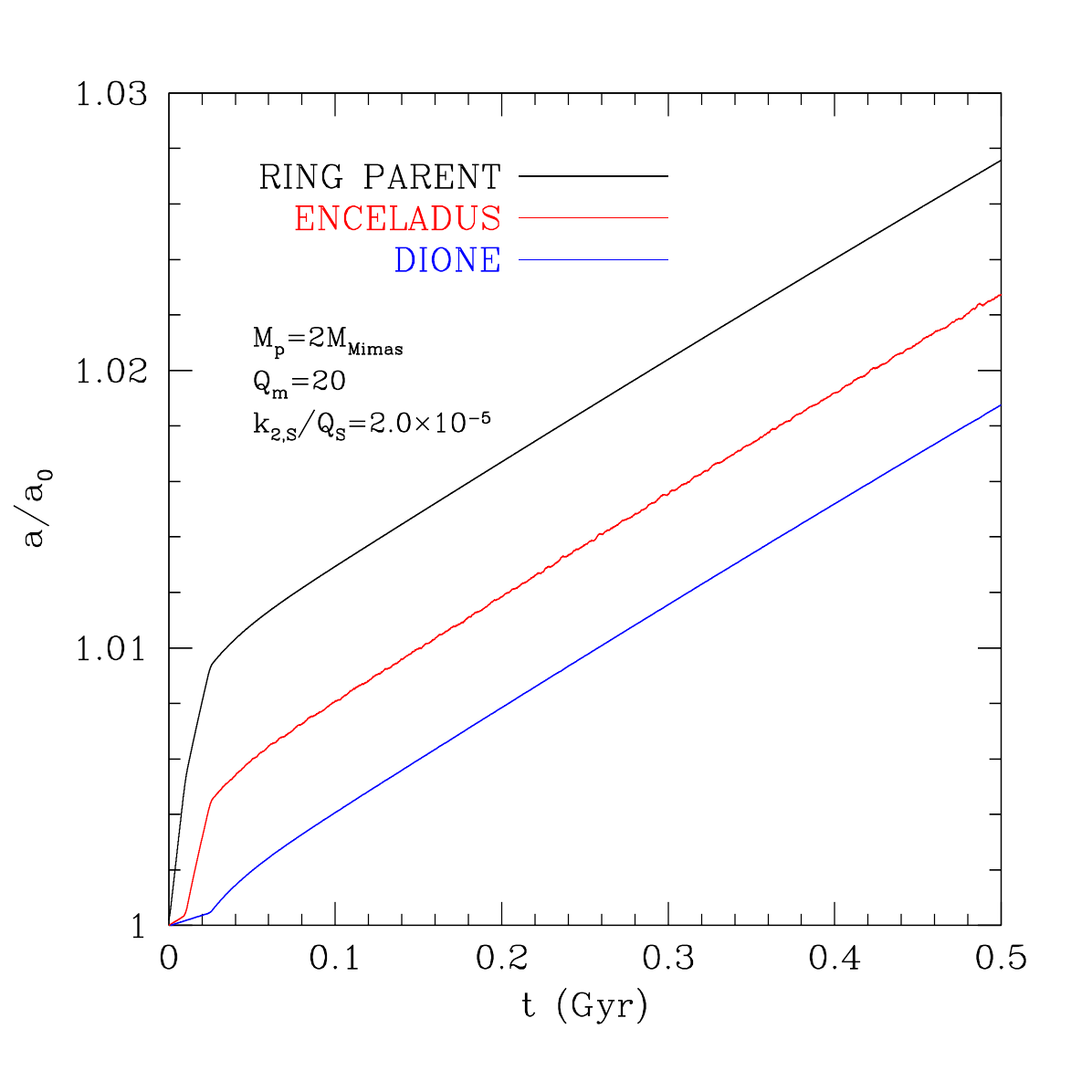}
\caption{The evolution of the semi-major axes of the ideal co-planar
system of 3 moons
consisting of the ring parent moon, Enceladus and Dione using
a conventional value of Saturn's tidal dissipation term.   Tidal
dissipation in Saturn causes the semi-major axis to grow.  The initial
semi-major axes are intentionally set to values slightly out of resonance
with values smaller than the current values.
The ring parent moon first locks into a 2:1 MMR with
Enceladus.  This system shortly after enters a MMR with
Dione that remains stable to the end of the integration of 0.5 Gyr.  This
calculation suggests that a ring parent moon could be trapped in mean
motion resonance near the Roche radius for a long time - perhaps as 
long as the age of the Solar system.}
\label{fig-a-conventional}
\end{figure}

\begin{figure}
\includegraphics[width=6.5in]{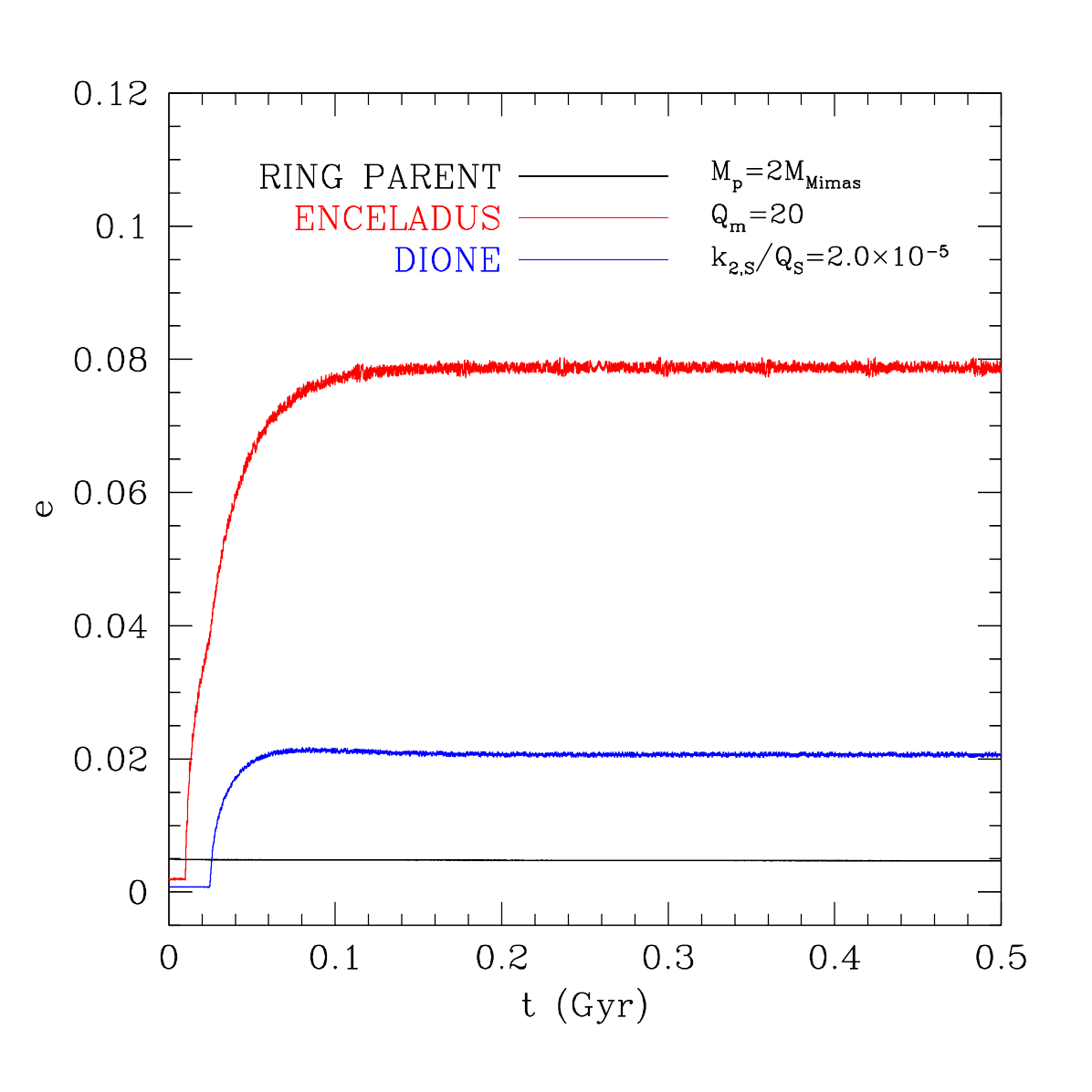}
\caption{The evolution of the eccentricities of the ideal co-planar system of 3 moons
consisting of the ring parent moon, Enceladus and Dione using a conventional
value of Saturn's tidal dissipation term.  When the moons
lock into mean motion resonance, the eccentricities of Enceladus and Dione
grow to equilibrium values of $e=0.08$ and $e=0.02$ respectively - values
an order of magnitude larger than the current values of $e=0.0047$ 
and $e=0.0022$.  The ring parent's eccentricity settles to $e=0.005$.  
These equilibrium values depend on the satellite Love numbers and
tidal dissipation terms $Q_m$.  Nevertheless, since the tidal heating is proportional
to $e^2$ \citep{peale99}, we expect internal heating rates about 2 orders
of magnitude larger than today in this scenario.  Prior to the removal of
the ring parent satellite, one therefore expects the icy mantle of Enceladus 
to be largely molten.}
\label{fig-ecc-conventional}
\end{figure}

Figures \ref{fig-a-conventional} and \ref{fig-ecc-conventional} 
show the evolution of the semi-major axes and eccentricities of the 3 moon system with a ring parent moon of twice the mass of Mimas as
assumed in the disruption simulations with the conventional value for Saturn's tidal dissipation.
Initially the semi-major axis of the ring parent grows rapidly in response to tidal friction with Saturn but the 3 moons quickly lock into MMR
slowing down the ring parent's outward migration significantly.  
The ring parent moon can therefore stay near the Roche zone for the age of the solar system and the 3 moon
system is stable for at least 0.5 Gyr with the conventional value of Saturn's tidal dissipation.
The resonance excites eccentricities in the 3 moons and equilibrium values for $e=0.005$, 0.08, and 0.02 are reached
for the ring parent, Enceladus and Dione respectively leading to heating rates from equation \ref{eq-peale} (in the same order) of $H=5$, 64, 37 GW.  The total heating rate is
slightly larger than the value estimated from equation \ref{eq-H} 
because of the significant eccentricities excited in the moons.  For the smaller ring parent
moon, the heating rate is sufficient to melt the ice and lead to a differentiated moon as assumed in the model.   The heating rate for Enceladus is very high by comparison and
implies the existence of an oceanic mantle and volcanically active past.  One can imagine significant mass loss during this phase which is consistent with the higher mean
density of $\rho=1.65\;{\rm g/cm^3}$ measured for Enceladus.   Even Dione with 10 times the mass of Enceladus would experience significant heating under these circumstances
implying a large subsurface ocean in the past and perhaps more surface geological activity than one might expect from the current state.   

\begin{figure}
\includegraphics[width=6.5in]{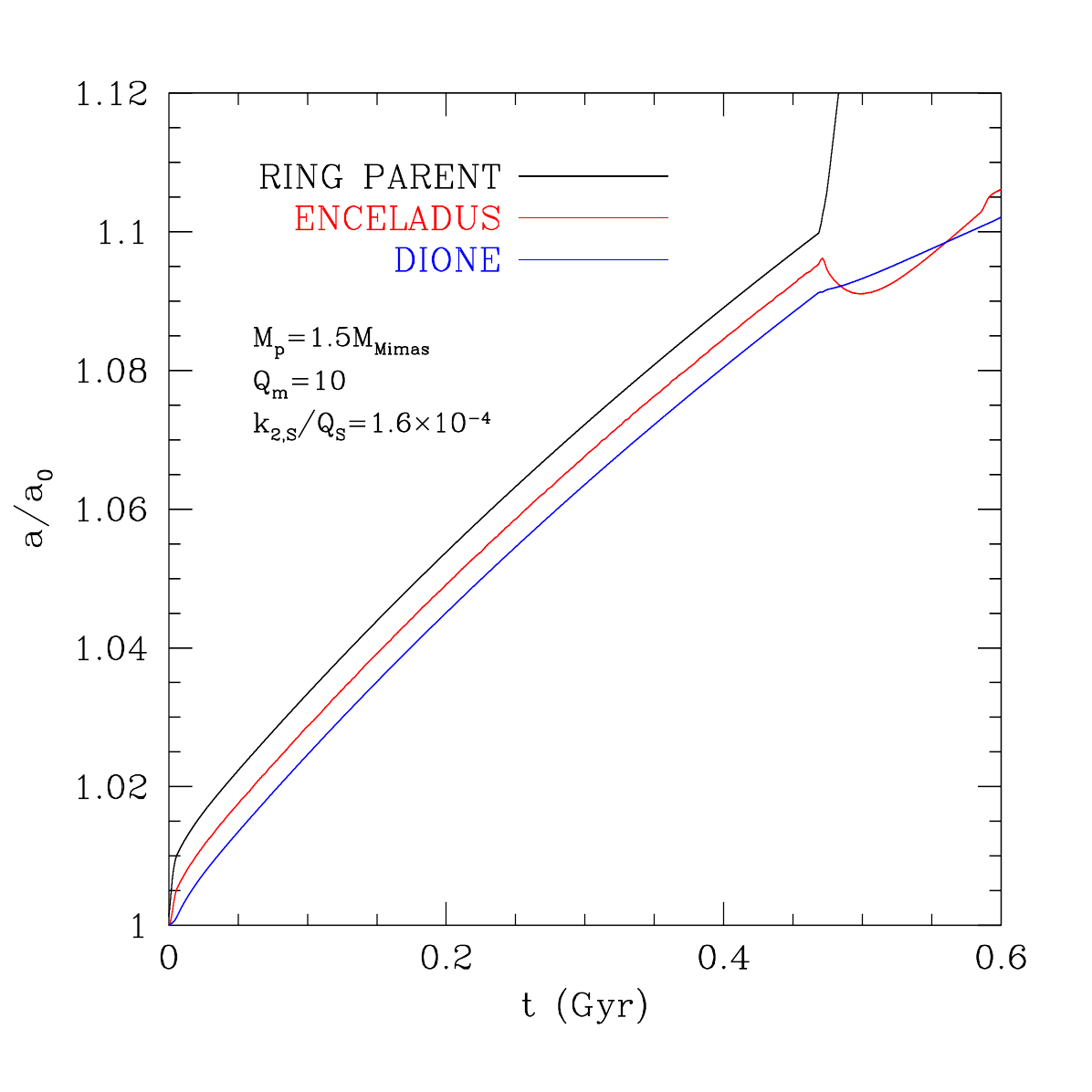}
\caption{The evolution of the semi-major axes of an ideal
system of 3 moons
consisting of the ring parent moon, Enceladus and Dione using
the \citet{lainey17} value of Saturn's tidal dissipation term.  The
ring parent moon has a mass $M_p=1.5\times M_{Mimas}$ and Love numbers
of the satellites are computed assuming $Q_m=10$.
This system only stays in MMR for about 0.5 Gyr before exiting the resonant state.}
\label{fig-a-lainey}
\end{figure}

\begin{figure}
\includegraphics[width=6.5in]{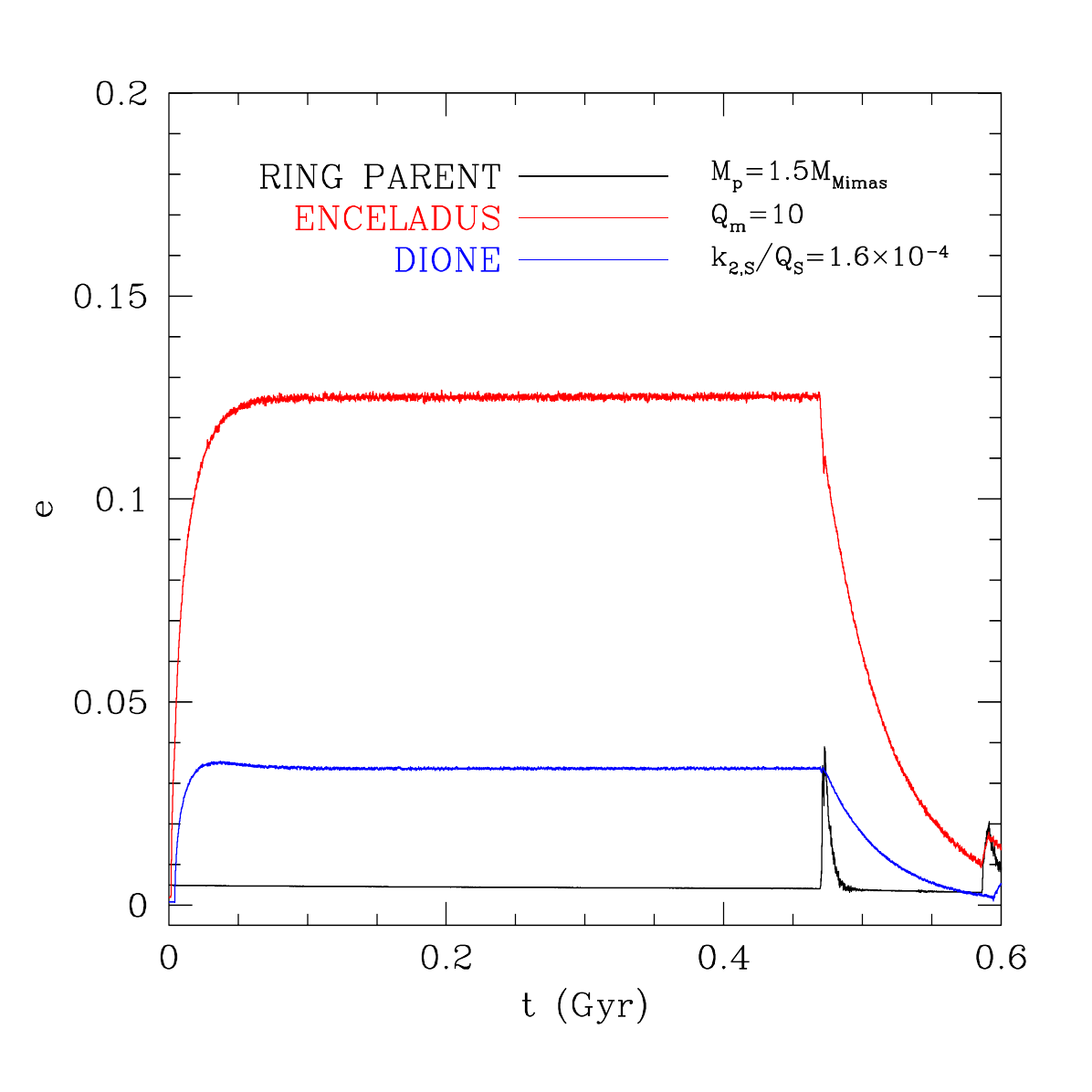}
\caption{The evolution of the eccentricities of an ideal system of 3 moons
consisting of the ring parent moon, Enceladus and Dione using the \citet{lainey17}
value of Saturn's tidal dissipation term.  When the moons
lock into MMR, the eccentricities of Enceladus and Dione
grow to equilibrium values of $e=0.125$ and $e=0.035$ respectively - values
that are larger than the fiducial case using a conventional dissipation term for Saturn.
The system abruptly exits MMR around 0.5 Gyr after the start of the integration and the
eccentricities decay rapidly.}
\label{fig-ecc-lainey}
\end{figure}

We repeated the calculations varying the mass of the ring parent moon and satellite Love numbers with Lainey's value for Saturn's tidal dissipation.  For the same ring parent mass with
$Q_m=20$ for the moons, the radial migration rate is 8
times faster as expected but the 3-moon system only stays in MMR for a short time.  
The failure of systems with larger ring parent moon masses and values of $Q_m$  to stay in MMR is probably due to the highly eccentric orbit excited in Enceladus.
Given that press reports suggest a smaller ring mass as determined by Cassini, we examined one case where
the ring parent moon is only $1.5\times$ the mass of Mimas.  With $Q_m=10$, this system remained in MMR for
approximately 0.5 Gyr (see Fig. \ref{fig-a-lainey} and Fig. \ref{fig-ecc-lainey}).  
The orbital radius of the ring parent grew by 10\%
from 135000 km to about 150000 km with the eccentricities of Enceladus and Dione excited to $e=0.125$ and 0.035 respectively resulting in significantly higher tidal heating rates.  
By the end of this time interval, the system falls out of MMR and the ring parent recommences a rapid outward radial migration.  
Lainey's tidal factor for Saturn implies that the system of icy satellites out to Dione may only have an age $\sim 1$ Gyr.   This calculation suggests
it may be possible to trap a ring parent satellite with appropriate properties for this timescale.  
A more thorough search of satellite parameters is needed to constrain this more precisely.

\begin{figure}
\includegraphics[width=6.5in]{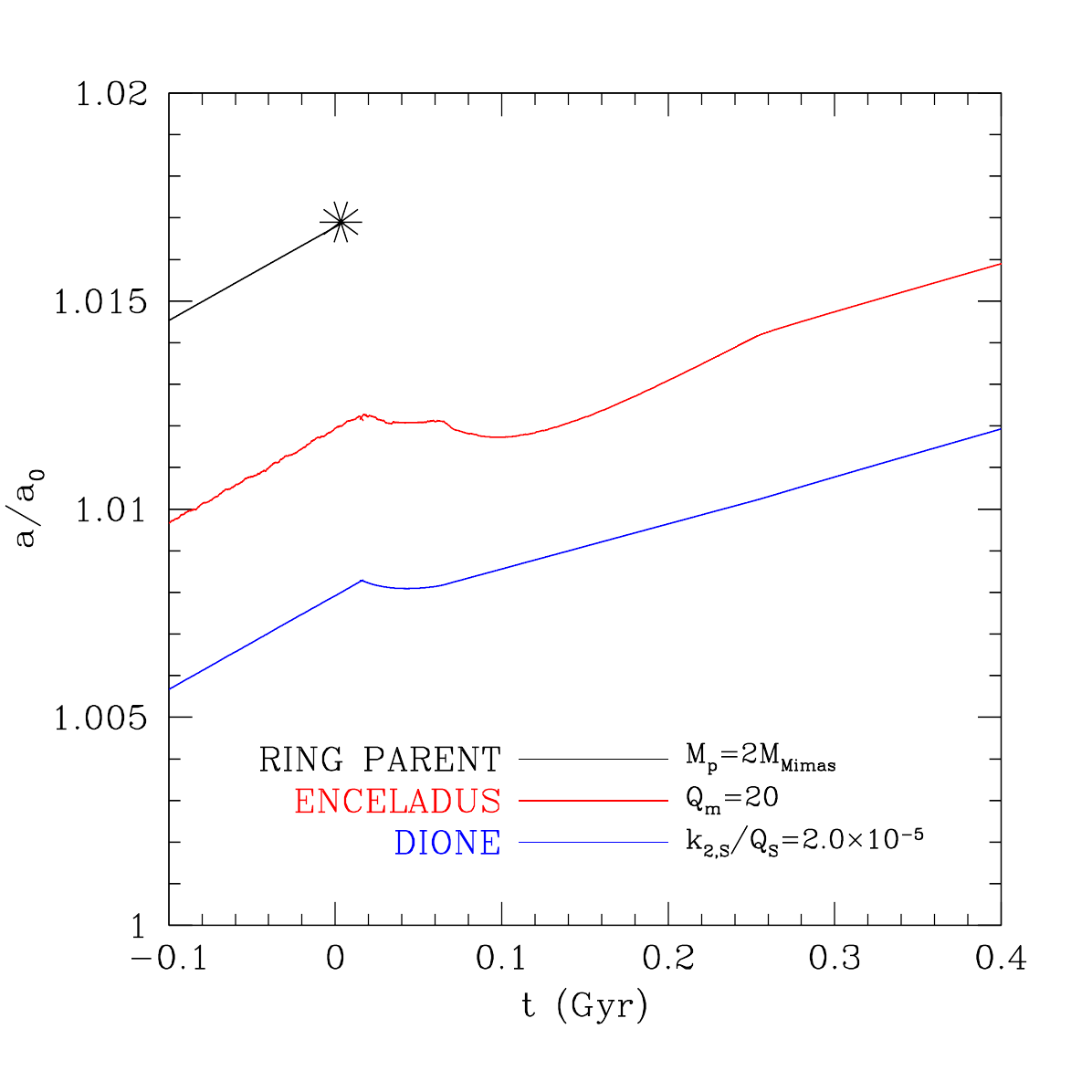}
\caption{The evolution of the semi-major axes of Enceladus and Dione
after the collisional disruption of the ring parent moon shown to
occur at $t=0$.  The tidal evolutionary growth of the semi-major axes
stalls for 0.1 Gyr but Enceladus and Dione re-enter a MMR
about 0.25 Gyr after the loss of the ring parent moon.}
\label{fig-a-postcollision}
\end{figure}

\begin{figure}
\includegraphics[width=6.5in]{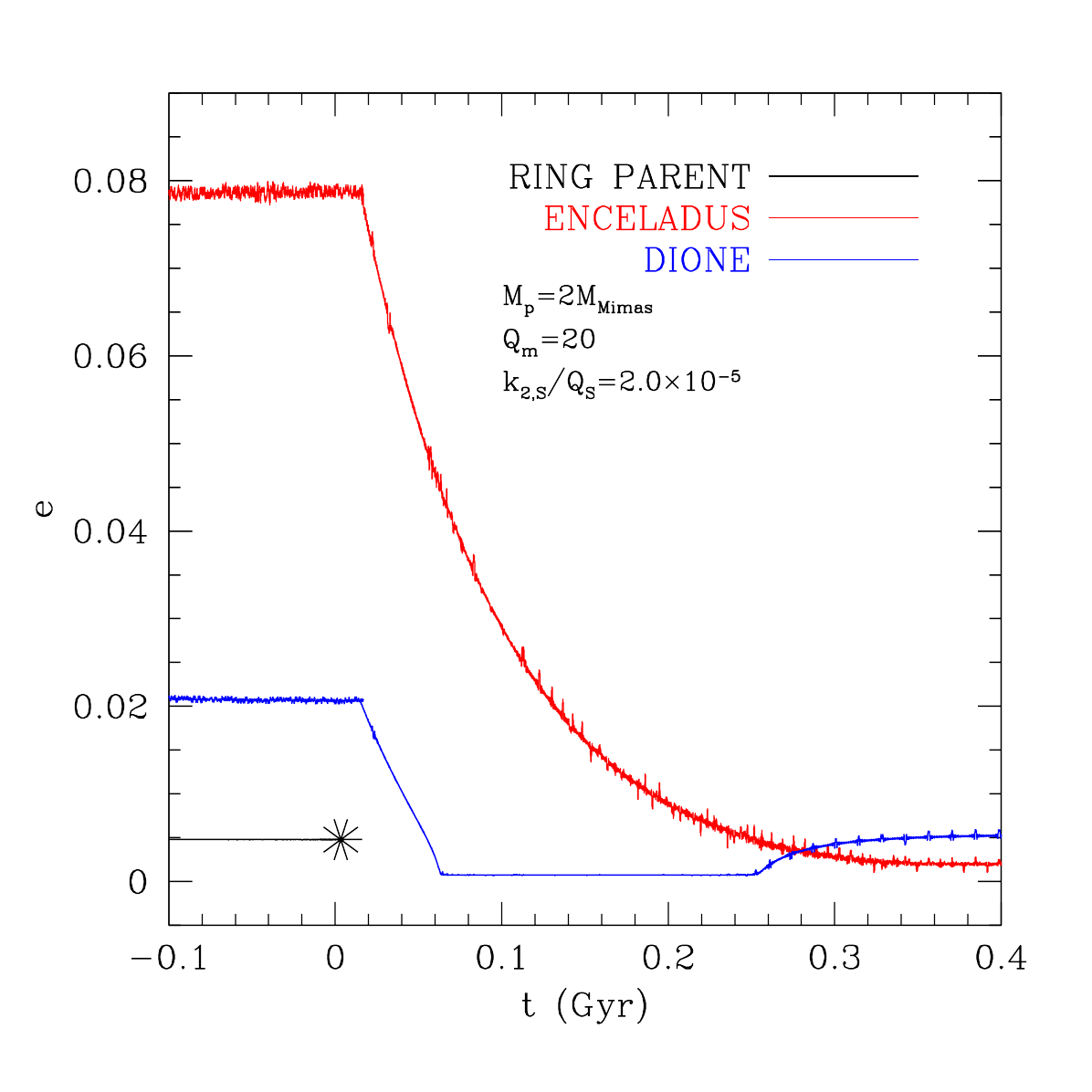}
\caption{The evolution of the eccentricities of Enceladus and Dione after
the collisional disruption of the ring parent moon at $t=0$.  The
eccentricities decay rapidly in the absence of the resonance but settle
into values comparable to the currently measured values.  While the choices
of the satellite Love numbers and tidal dissipation functions $Q$ are
only estimates, in this scenario Enceladus and Dione enter the current
resonance and orbital configuration about 0.25 Gyr after the collision.
This is close to the timescale one expects for a newly formed Mimas to
reach its current orbital radius after accretion of the ring debris onto the collision remnant.}
\label{fig-ecc-postcollision}
\end{figure}

Finally, we examined the orbital evolution of Enceladus and Dione after the disruption of the satellite for the case using the conventional value of tidal dissipation
and ignoring ring torques.
We simply remove the ring parent moon from the system in MMR with
equilibrium eccentricities and continue the orbital integration for the two remaining moons.  
Figures \ref{fig-a-postcollision} and \ref{fig-ecc-postcollision} show the evolution of the semi-major axes
and eccentricities after the impact.  The tidal migration of Enceladus and Dione stalls for a few hundred million years but the two moons eventually settle back into a 2:1 MMR.
The eccentricities decay with the timescales expected when resonant forcing is removed and a small eccentricity is re-excited in Dione when they get back into resonance.   
We note that this illustrative example is not identical to the current MMR between Enceladus and Dione which instead excites an eccentricity in Enceladus 
rather than Dione.
There are in fact a multiplet of possible 2:1 resonances that can excite both inclinations and eccentricities of 
either Enceladus or Dione or both \citep{sinclair72,zhang09}. 
The convergent co-planar orbital scenario 
illustrated here passes through the $e$-Dione resonance first and 
the moons are apparently captured here contrary to the current state of the system.
\citet{zhang09} have argued that capture into 
the $e$-Dione resonance becomes less probable if Dione has a prior free eccentricity greater than a critical value $e\sim 0.002$.  A more
thorough examination of the orbital evolution in this scenario with varying values of $k_2/Q_m$ for the moons might 
help understand different pathways to the current state.
In any case, the timescale
for the eccentricity of Enceladus to decay to its present value is approximately 0.3 Gyr which coincides with 
the expected timescale for the newly formed Mimas to migrate to its
orbital radius.  One still expects remnant heat from the previous highly eccentric orbit of Enceladus and this 
may explain why it is still partially molten and quite active despite the low
amount of tidal heating at the present time.

\section{Conclusions}

In summary, the collisional disruption mechanism proposed in this paper presents a solution to the origin of
Saturn's rings that
is consistent with the mass, age and composition of the rings.    A differentiated moon
with a mass twice that of Mimas and an ice fraction consistent with the other icy moons located near the 
Roche radius can be disrupted by a heliocentric comet with a radius of $\sim 20$ km.  Representative simulations
show that the icy mantle can be liberated in the collision forming a nearly pure ice ring system with a mass comparable
to Mimas while leaving behind a rocky remnant.  It is argued that the viscous spreading of the ring of debris will lead to the current
ring system and a new satellite composed of the remnant and re-accreted debris that we identify with Mimas.
While this event could have occurred any point during the Saturn system's lifetime, the short timescales for viscous spreading of the rings
and contamination from meteoroid bombardment suggest that the event happened within the last few hundred million years.  
The probability that an impact with sufficient destructive energy occurs in this timeframe is $\sim 0.01$ making it a $2-3\sigma$ event -
perhaps unlikely but not wildly so \citep{zahnle03,dones09}.
The model therefore simultaneously accounts for a mass of the rings comparable to Mimas, a young age and nearly pure ice composition.
A further requirement for this model to work is the necessity that the ring parent moon is locked in MMR with Enceladus
and Dione so that it remains trapped near the Roche radius prior to destruction.   We have demonstrated that this phase would 
tidally heat Enceladus at a much greater rate than now.   Remnant heat might still be present explaining the
existence of Enceladus' current heat output, subsurface ocean and surface activity.

During Cassini's Grand Finale tour in 2017, the spacecraft flew close enough to the rings to measure a 
gravitational perturbation using the radio science experiment 
and so determine a definitive mass for the rings \citep{charnoz18}.
The scenario proposed here only works if the ring mass is comparable to the mass of Mimas.
If the mass of the rings is close to the upper range of current estimates - about $3\times$ 
the mass of Mimas - this scenario seems less likely since it becomes increasingly difficult
to keep such a large ring parent moon within the Roche zone because of the stronger torques from tidal friction
and the lower probability for disruption by an impact. 
New as yet unpublished reports suggest that the total mass of rings may be less than half of the mass of Mimas.
Once the mass is known precisely,
this scenario is strongly constrained and it should be possible to narrow 
the range of collision energies, impact geometries and the initial composition of the ring parent moon using more
sophisticated simulation methods.

%\begin{table}
%	\centering
%	\caption{This is an example table. Captions appear above each table.
%	Remember to define the quantities, symbols and units used.}
%	\label{tab:example_table}
%	\begin{tabular}{lccr} % four columns, alignment for each
%		\hline
%		A & B & C & D\\
%		\hline
%		1 & 2 & 3 & 4\\
%		2 & 4 & 6 & 8\\
%		3 & 5 & 7 & 9\\
%		\hline
%	\end{tabular}
%\end{table}
%

\section*{Acknowledgments}

I acknowledge Peter Goldreich, Chris Thompson, Yanqin Wu and Luke Dones
for discussions that have inspired and clarified points in this paper.
I thank referees Luke Dones and Matija {\'C}uk for comments that pointed
to pertinent observations and helped
elucidate important physical processes as well
as point to many useful references.
I also thank Bj{\"o}rn J{\'o}nsson for permission to use his texture
map of Saturn from Cassini data for Figure \ref{fig-saturnseq} and the accompanying 
animation.  Simulations were carried out at the computing facilities
of the Canadian Institute for Theoretical Astrophysics.  This research 
was financially supported by the Natural Sciences and Engineering Research
Council of Canada.

%%%%%%%%%%%%%%%%%%%%%%%%%%%%%%%%%%%%%%%%%%%%%%%%%%



%\bibliographystyle{mnras}
%\bibliographystyle{elsarticle-harv}
%\bibliographystyle{model2-names.bst}\biboptions{authoryear}
\bibliography{refs}

%\begin{thebibliography}{99}
%\bibitem[\protect\citepauthoryear{Author}{2012}]{Author2012}
%\bibitem[\protect\citepauthoryear{Others}{2013}]{Others2013}
%\end{thebibliography}

%%%%%%%%%%%%%%%%%%%%%%%%%%%%%%%%%%%%%%%%%%%%%%%%%%

%%%%%%%%%%%%%%%%% APPENDICES %%%%%%%%%%%%%%%%%%%%%

\appendix
\section*{Appendix A: Collisional N-body Algorithm}
\setcounter{equation}{0}
\renewcommand\theequation{A\arabic{equation}}
\setcounter{figure}{0}
\renewcommand\thefigure{A\arabic{figure}}

Each particle in a system is represented as a hard massive sphere with a finite radius and 
frictionless surface with no internal angular momentum.
The particles move under the influence  of gravity and hard collisions with the equations of motion 
integrated using a second order leapfrog method.
Gravitational forces for the system are computed using a parallelized treecode algorithm 
\citep{barnes86,dubinski96}.  At each fixed step, the computed
acceleration on each particle is used to change the particle velocity.   
At the step midpoint, a particle drifts at its current velocity and a
nearest-neighbour finding algorithm is used to locate all particles within a fixed search radius 
that could possibly collide with that particle.  All
possible collisions are determined over the time interval and sorted in time.  
The particles are then moved forward in position to the time of the first collision in
the sorted list.  
The velocities for the first pair of colliding particles (now touching) are adjusted by 
reversing the radial velocities thus conserving linear momentum through the collision.  We assume the
spheres have a frictionless surface so there is no transfer of angular momentum.  
After the first collision is complete, the neighborhood of this pair of
particles is searched again for any new possible collisions.   
If new collisions are detected, they are added to the chronologically sorted list and the
possible cascade of collisions caused by this first collision are detected and used to modify 
and re-sort the collision list.  Once this is done, one
then proceeds to the first collision in this new list and repeats the procedure again with the 
modified collision list.  
The sequence of collisions is therefore followed in order to the end 
of the time interval.  
Once the collision sequence is completed the particles are moved to their final position 
at the beginning of the next time interval, gravitational accelerations are computed again, 
velocities are updated and the entire procedure is repeated through the next timestep.

The algorithm is parallelized in the following way.  The treecode method uses the concept of 
locally essential trees \citep{dubinski96}.  In this method, the
system is divided into rectangular subdomains in 3 dimensions each containing a unique subset 
of particles in the system mapped to a processor on a
parallel supercomputer.  A Barnes-Hut (1986) oct-tree structure is built for each subdomain used to compute 
local gravitational forces.  To account for the external
gravitational forces on particles from exterior subdomains, 
the relevant pieces of tree structures are imported to create a locally essential
tree that contains enough information to compute the gravitational forces to the accuracy prescribed 
by the treecode algorithm.   

The locally essential tree is also leveraged by the nearest-neighbour finding algorithm used for 
collision detection.   Around a chosen particle,  the tree
structure is used to search the entire list of particles to find the nearest subset to test 
for possible collisions.   It is hopelessly inefficient to search the
entire list, so to speed up the process only the N nearest particles are examined 
where $N\approx 27$ typically.   The oct-tree structure used to compute
gravitational forces can be used efficiently to generate this list for each particle in a domain.  
The locally essential tree also contains particles from
neighboring domains and so the procedure described above for determining and moving through a list 
of collisions can be applied in parallel in each subdomain
with the small penalty of computing some collisions more than once for particles near the boundary.  
We find in practice this algorithm works well in
parallel with as many as 1024 processors for the systems containing $10^{7-8}$ particles.   

We incorporate collisional energy dissipation by incorporating inelasticity through a coefficient of restitution $\epsilon<1$.
For inelastic collisions, the rebound velocity is reduced by a factor $\epsilon$ following other algorithms \citep[e.g.,][]{richardson00}.
In practice, for the hypervelocity impacts simulated in this paper we assume the collisions are elastic with $\epsilon=1$
during the impact and disruption phase.  When the debris from the collision disperses to form a ring, we use $\epsilon=0.9$
to dissipate energy.

With this algorithm, the possibility exists for missing some collisions of particles 
located near the domain boundaries.  
This pathology can lead to some errors in collision detection.   A small fraction of particles can end up
overlapping with their relative distance being less than a particle diameter.  
Any erroneous overlapping particles at the end of a
time interval are assumed to be in a state of collision and their radial velocities are reversed. 
In practice, less than 0.1\% of particles are overlapping between timesteps and
if this problem arises the remedy is to reduce the time interval between steps until the fraction 
of overlappers is reduced to an acceptable level.  

The collisional N-body code is available to other researchers by request.

%that depends on the relative velocity of particles:
%\begin{equation}
%\epsilon_v = 1 - (1-\epsilon_0)T(v)
%\end{equation}
%where $\epsilon_0=0.9$ and $T(v)$ is a step-like function:
%\begin{equation}
%T(v) = (1 + e^{-(v-v_0)/\delta v})^{-1}
%\end{equation}

\section*{Appendix B: Initial Conditions for Gravitating Rubble Piles}
\setcounter{equation}{0}
\renewcommand\theequation{B\arabic{equation}}
\setcounter{figure}{0}
\renewcommand\thefigure{B\arabic{figure}}

In this appendix, we describe methods for setting up models of differentiated
moons described as rubble piles composed of hard spheres of different densities to represent rock and ice.
In condensed matter physics,  systems of colliding hard spheres are used
as idealized models for dense gases and liquids and can 
be described by a well-defined equation of state (EOS).   
The hard-sphere EOS depends on the packing fraction defined by
$\eta = 4\pi/3 r_0^3 n$ with $n = N/V$ is the number density 
and $r_0$ is the radius of the hard spheres.   
The maximum value of $\eta$ is
thought to be close to the maximum value of a system of hexagonally close-packed (HCP) spheres 
with a value $\eta_{max}=\pi\sqrt{2}/6\approx 0.74$. 
%but
%for a randomly close-packed (RCP) system the value is estimated 
%empirically as $\eta\approx 0.64$ (REF).   
The EOS is defined through the
compressibility $Z\equiv PV/NkT$:
with $Z$ being some function of the packing fraction $\eta$.  
The compressibility can be expressed as a power series in $\eta$:
\begin{equation}
Z = 1 + \sum_{k=2}^{n_{max}} b_k \eta^{k-1}
\label{eq-virial}
\end{equation}
where the $b_k$ are the virial coefficients, generally determined empirically 
through numerical simulations of hard sphere systems \citep[e.g.,][]{clisby06}.  
Simpler analytic expressions, that are accurate to high order, exist such as
the Carnahan-Starling (CS) EOS compressibility \citep{carnahan69} is:
\begin{equation}
Z = \frac{1 + \eta + \eta^2 - \eta^3}{(1 - \eta)^3}.
\end{equation}
We rewrite the equation of state in the form $P = \rho \sigma^2 f(\rho)$ 
based on the definition of $\eta$ and $Z$ where $\rho$ is the density and
$\sigma^2 = kT/m$ is the velocity dispersion.
There is a phase transition between $\eta=0.494$ and $\eta=0.545$ with approximately fixed $Z\approx 11.5$
where the hard-sphere systems freezes to a crystalline solid when using hexagonal close-packing.  
For $\eta < 0.494$, we use Z as defined in equation \ref{eq-virial}.
For $\eta > 0.545$,
the compressibility is accurately described by the empirical formula:
\begin{equation}
Z = \frac{3}{1-z} - \frac{a(z-b)}{z-c}
\label{eq-speedy}
\end{equation}
where $z = \eta/\eta_{max}$ and $a=0.5935$, $b=0.708$ and $c=0.601$ as calculated by \citet{speedy98}.

To determine spherical equilibrium solutions for a self-gravitating system of hard spheres,
we need only solve the equations of hydrostatic equilibrium
%\begin{eqnarray}
%\nabla P &=& -\rho \nabla \Phi \\
%\nabla^2 \Phi &=& 4\pi G \rho
%\end{eqnarray}
to determine a density profile for a given initial central density for the rubble pile.  
Since there are no internal sources of energy,  the solutions are isothermal.   We therefore
assume a Maxwell-Boltzmann velocity dispersion $\sigma^2$ for the system of particles.   
The hydrostatic equations can be solved by choosing a central density
with a guess for the velocity dispersion and integrating the equations to determine 
a density profile and total mass.  
The correct velocity dispersion can be determined by iteratively adjusting its value up or down depending on whether the total
mass of a solution is less or greater than the desired value.  We isolate the correct value of the velocity dispersion using a binary search.
The end result is a spherical density profile and a single velocity dispersion.  To set up the initial conditions for particle
positions, we first lay out particles in hexagonally close-packed lattice of uniform density 
and carve out a sphere containing the
desired number of particles.  Given the mass profile $M(r)$ derived above we then adjust the particles radial positions according
to $r \propto M^{1/3}$ so that radial mass profile is matched.  Particle velocities are sampled from a Maxwell-Boltzmann distribution
with the given velocity dispersion.  

For a system with a single particle density, this procedure results in collisional hard sphere system in hydrostatic equilibrium.
To model a differentiated moon, we apply a simple modification to the above procedure to set up a model with a rocky core and icy
mantle.  When initializing the unperturbed close-packed lattice of particles, we simply assign a higher mass to particles
in the core according to the ratio of densities of rock to ice (approximately $3\times$).  The computation of the mass profile in
hydrostatic equilibrium requires an adjustment to account for the greater particle density in the center but this is
straightforward to implement within the iterative procedure.  There is also a discontinuous change in the velocity dispersion of 
particles when moving from the rock core to the ice mantle to compute virial equilibrium.
For systems with densities near maximum close-packing there
can be discontinuities in the density profile within the core which are apparent in our models though the jump in density through
the discontinuity is generally small.

\begin{figure}
\includegraphics[width=6.5in]{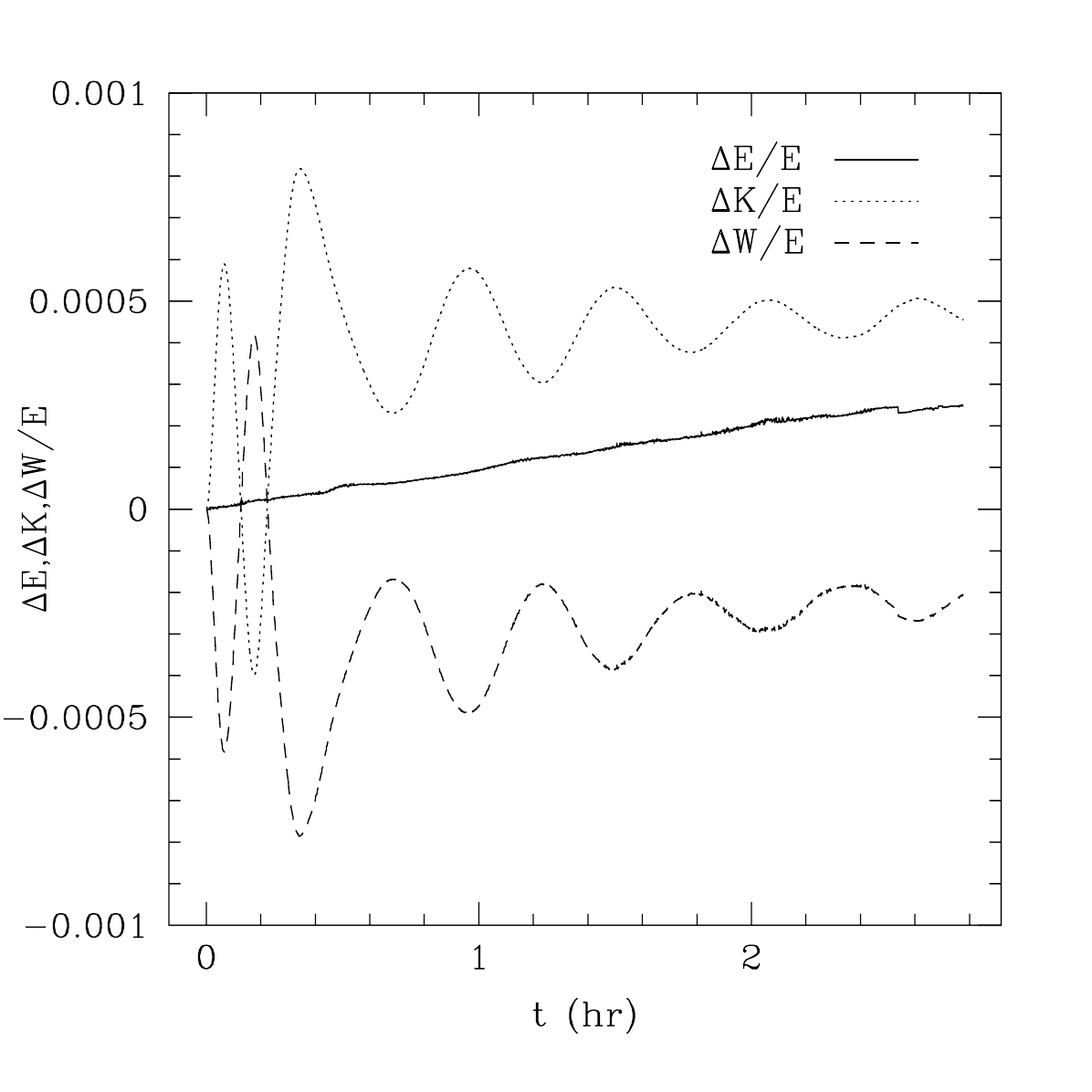}
\caption{Energy evolution towards equilibrium for the 
initial spherical moon model.  The system is evolved dynamically using the N-body code for several dynamical
times.   For a system in equilibrium, the total kinetic and potential energies should be constant in time.
The system pulsates with energy being exchanged between kinetic and potential forms at the 0.1\% level but the pulsations damp away after
a few hours.   The total energy of the system drifts by 0.02\%.  The systems are almost in perfect virial equilibrium validating the input equation of state and the
collisional N-body method.}
\label{fig-e}
\end{figure}

\begin{figure}
\includegraphics[width=6.5in]{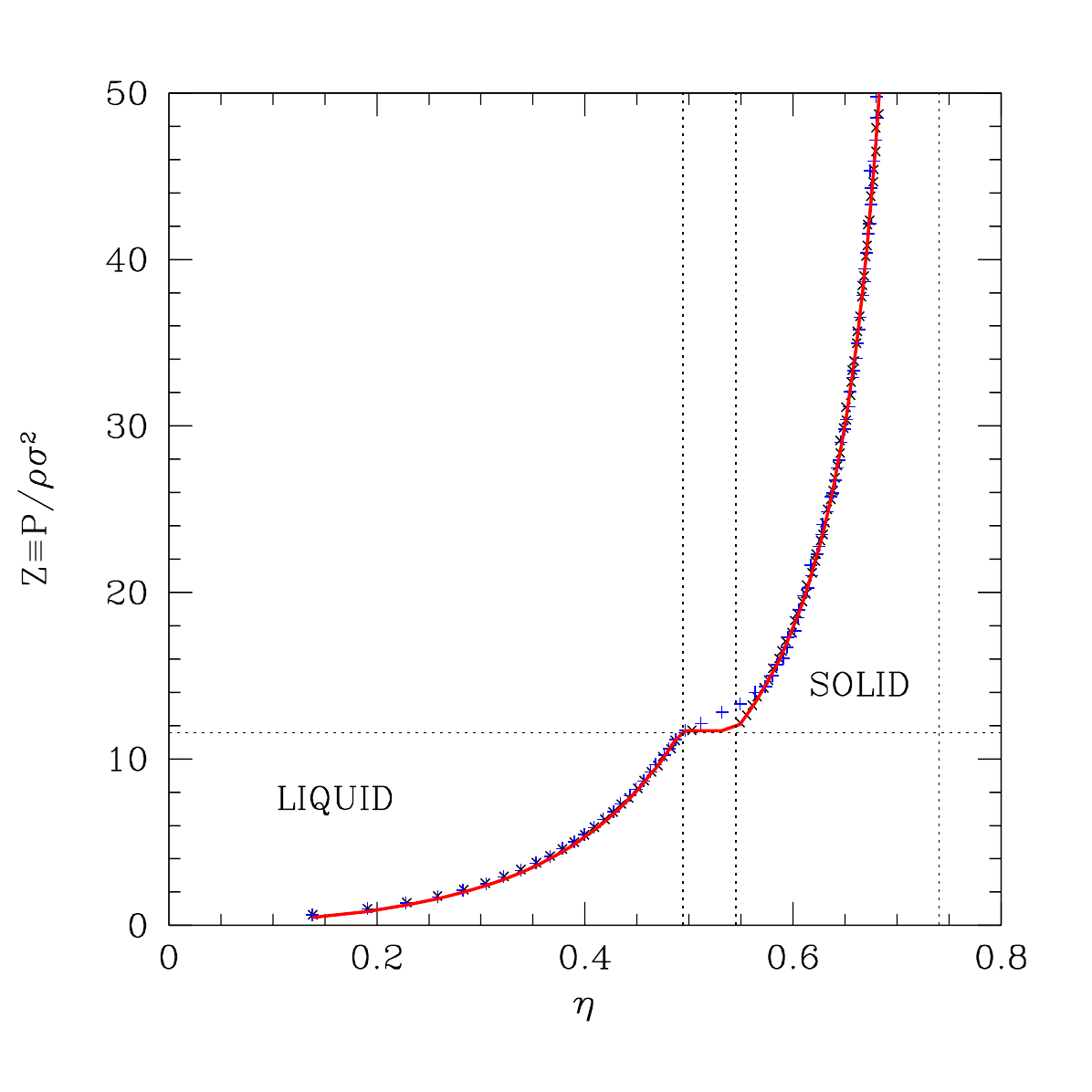}
\label{fig-Z}
\caption{Compressibility measured from the equilibrium model at the initial (black crosses) and final times (blue crosses) compared with the theoretical
expectation (red line).  The agreement is very close and shows that both the equation of state and methods for computing the initial hydrostatic
profile are consistent.}
\end{figure}

We dynamically evolved the initial model to
validate the equilibrium state. 
The density profile for the model moon used in this paper has been shown previously in Figure \ref{fig-den}.    The main points to note are the smooth drop off
to zero density at the periphery compared with the ideal homogeneous model and the small discontinuity in density at $R=240\;{\rm km}$ due to the phase
change.  Figure \ref{fig-e} shows the relative change in potential, kinetic and binding energy for the model as a function of time for
the $N=10M$ particle model.  The model is slightly out of equilibrium with potential and kinetic energy being exchanged with amplitudes of 0.1\% of the total binding
energy.   The system pulsates for several oscillations over the course of few hours but eventually damps down.   The
total binding energy in principle is a conserved quantity but integration errors cause it to drift by 0.01\%.  
In summary, using the hard-sphere EOS provides a useful away to set up rubble-pile models in hydrostatic equilibrium.
We also plot the compressibility
$Z=P/\rho\sigma^2$ as a function of the packing fraction measured from the N-body model (Figure \ref{fig-Z}).  The agreement between the measured and
theoretical compressibility is very close showing the consistency of the equation of state with hydrostatic equilibrium.

The initial conditions code used to set up model moons in hydrostatic equilibrium  is available to other researchers by request to the author.



%%%%%%%%%%%%%%%%%%%%%%%%%%%%%%%%%%%%%%%%%%%%%%%%%%

% Don't change these lines
%\bsp	% typesetting comment
%\label{lastpage}
\end{document}